\documentclass[%
 reprint,
 amsmath,amssymb,
 aps,
]{revtex4-2}

\usepackage{graphicx}
\usepackage{dcolumn}
\usepackage{bm}%
\usepackage{amsmath}
\usepackage{mathtools}
\usepackage{braket}
\usepackage{physics}
\usepackage{afterpage}
\usepackage{nicefrac}
\usepackage[svgnames]{xcolor}
\usepackage{amsfonts}
\usepackage{amssymb}
\usepackage{comment}
\usepackage{stfloats}
\usepackage{float}
\usepackage{subfigure}
\usepackage[subfigure]{tocloft}

\newcommand{\EatOneArg}[1]{}

\begin{document}

\preprint{APS/123-QED}

\title{Entanglement transfer of a Rydberg W-state to a multi-mode photonic state}%

\author{Aneesh Ramaswamy}
\affiliation{Stevens Institute of Technology, Hoboken, NJ, USA}

\author{Svetlana A. Malinovskaya}%
\email{smalinov@stevens.edu}
\affiliation{Stevens Institute of Technology, Hoboken, NJ, USA}

\date{\today}

\begin{abstract}
A robust quantum protocol has been developed  that achieves highly efficient entanglement transfer from a three-atom Rydberg system,  initially in a W state $(\ket{rrg}+\ket{rgr}+\ket{grr})/\sqrt{3}$, to an equivalent photonic W state $(\ket{101}+\ket{110}+\ket{011})/\sqrt{3}$. The 
entanglement transfer is achieved by dynamically adjusting the cavity mode frequencies and modulating the coupling rates, simplifying the complex transfer process into a sequence of processes involving two-level avoided crossings. 
We demonstrate that entanglement transfer can be achieved using either a fully adiabatic protocol or one with controlled non-adiabatic transitions at avoided crossings, generated by continuously chirping the cavity modes. Our adiabatic protocol uses the fractional STIRAP method to facilitate the partial population transfer required for generation of the photonic W state. In comparison, the non-adiabatic protocol uses non-adiabatic transitions to achieve the required partial population transfer. 
Furthermore, we propose two strategies for experimental implementation of our protocols.
\end{abstract}

\maketitle

\section{Introduction}

In the past decade, significant strides have been made in both theoretical and experimental exploration of the distinctive quantum mechanical phenomenon of entanglement, bearing profound implications for the advancement of scalable technologies for  quantum information \cite{Yu2021, Erhard2020}. 
Efforts to produce entangled states on a large scale have yielded promising outcomes across multiple platforms, including photons and neutral atoms. Notably, photons hold particular significance for quantum communications and the realization of a photonic quantum network \cite{Covey2023}. On the other hand, 
ultracold Rydberg atoms offer a versatile platform for the generation and manipulation of entangled states of atoms 
\cite{Saffman2010, Lee2016}, and photons \cite{Ye2023, Ghosh2021}. 
Their distinctive characteristics, including robust long-range interactions and remarkably prolonged lifetimes, position Rydberg atom arrays as promising candidates for scalable quantum computing.

Various platforms have been proposed to realize a scalable quantum computing paradigm, including trapped ions \cite{Cirac1995, Bruzewicz2019}, neutral atoms \cite{Henriet2020, Covey2023}, superconducting quantum interference devices (SQUIDs) \cite{Wendin2017, Huang2020}, and photons \cite{Takeda2019}. 
While photons exhibit promising features like weak environmental interactions and efficient information transfer capabilities, they have not progressed as rapidly as other platforms in terms of error correction and fault tolerance. In contrast, trapped ions and SQUIDs have emerged as dominant platforms for realizing fault-tolerant quantum computing. 
However, recent advancements in neutral atom arrays using Rydberg atoms have significantly propelled this platform as a promising candidate. This progress stems from achieving a logical quantum processor with dynamically programmable arrays and coherent transport facilitated by optical tweezers \cite{Bluvstein2022, Bluvstein2023}. Consequently, there is now a renewed emphasis on investigating Rydberg atom dynamics for realizing fast and robust quantum gates through holonomic operations \cite{Zhang2023, Liu2020}, for generating various classes of entangled states like the GHZ and W states \cite{Pachniak2021, Haase2022}, for exploring quantum memory applications, and for transferring entanglement to other systems such as cavity photonic states. The W state, in particular, plays a crucial role in addressing specific quantum computational challenges, such as leader election in anonymous quantum networks \cite{DHondt2006}. Our objective is to produce $n$-photonic W states in a multimode cavity by transferring entanglement from an equivalent $n$-atom W state. 
The process of transferring entanglement to cavity photons contrasts with the achievements made in producing Rydberg entangled states, which were showcased in our past research where we demonstrated the capability to generate GHZ and W Rydberg states through the employment of chirped light pulses \cite{Pachniak2021}. In that work, chirped STIRAP \cite{Vitanov1999, Vasilev2008} was employed to produce GHZ states, while nonadiabatic transitions were utilised to create W states. In this work, we show that we can generate the photonic W state with and without the use of non-adiabatic transitions. The findings presented in this work will contribute to the advancement of robust hybrid quantum gates tailored for quantum networking applications \cite{Liu2022}.

We have formulated multi-stage protocols for transferring populations from multi-atom states to multimode photonic states, focusing on the system involving a three-atom W state and a three-cavity mode state. To achieve efficient transfer, we have developed an adiabatic protocol based on the F-STIRAP method with chirp \cite{ChathanathilPRA2023} 
which required modulating the cavity coupling rates to emulate Gaussian pulses and applying linear chirping to the frequencies of the cavity modes. We have also developed an alternate non-adiabatic  protocol where we adapted the rapid adiabatic passage (RAP) technique \cite{RAP}.
To attain the necessary modulation of the coupling rates and mode frequencies, we proposed a strategy for implementation using a 3D rectangular cavity with a moving mirror. 
An alternative strategy involving the use of multiple chirped pulses of radiation and cavity modes has also been proposed. In this strategy, the transition from the ground to the Rydberg atomic state is mediated by a two-photon process involving the cavity coupling and the chirped pulses, with each driving a one-photon transition to an off-resonantly excited intermediate state.
Other methods in the literature to generate the Gaussian modulation of the coupling rates required for STIRAP-based protocols include propagating an atom with constant velocity across the cross section of an optical cavity with Hermite-Gauss modes \cite{AMNIATTAL}.

In order to develop robust quantum protocols for entanglement transfer, we employed a series of transformations to separate our total system into a sum of decoupled super-effective two-level systems (TLSs), greatly simplifying the dynamics occurring throughout the protocol. This allowed us to derive analytic results and numerics, even in the case of our non-adiabatic protocol where we used the adiabatic impulse approximation \cite{Damski2006}. Ultimately, we demonstrated that our protocols successfully generate the photonic W state, as required. In the subsequent section, we commence by formulating a detailed problem statement, followed by a comprehensive derivation of the solution leading to the development of our quantum protocol.

\section{Problem statement}\label{Sec5Prob}
We consider a system of three atoms trapped in an optical lattice, which is inserted into a three-mode cavity.
The goal is to transfer the entanglement from atoms to photons to generate a three-mode entangled photonic state, the target state. Initially, the atoms are prepared in a superposition of the ground and Rydberg states forming a W state and the multimode cavity is in the vacuum state. The initial state reads, 

\begin{equation}
	\Psi_0=\dfrac{1}{\sqrt{3}}\left(\ket{rrg}+\ket{rgr}+\ket{grr}\right)\otimes\ket{000}.\label{Psii}
\end{equation}

The target state reads, 
\begin{equation}
	\Psi_1=\ket{ggg}\otimes\dfrac{1}{\sqrt{3}}\left(\ket{110}+\ket{101}+\ket{011}\right),\label{Psif}
\end{equation}
with accuracy up to a global phase. 
The Rydberg-Rydberg inter-atomic repulsion differentiates the energies of Rydberg nearest-neighbour states $\ket{rrg}, \ket{grr}$ from those of the next-nearest neighbor state $\ket{rgr}$ and the single Rydberg atom states $\ket{rgg}, \ket{grg}$, and $\ket{ggr}$. 

Consequently, we can map each cavity mode to specific transitions between three-atomic states  
with the atoms being in one of the aforementioned three-atomic configurations or in the all-ground state $\ket{ggg}$. Each transition frequency remains sufficiently off-resonant with transitions corresponding to the other configurations.
Therefore, we can assign each cavity mode to a single transition frequency. However, an immediate problem arises. Every path to the $\ket{ggg}$ state involves a transition from a state with a single Rydberg atom, e.g. $\ket{ggr}$, meaning that the atom-cavity state $\ket{ggg}\ket{011}$ 
cannot be realized. A way to circumvent this problem is to force the first mode to be off resonance with this transition while the second mode is simultaneously in resonance for a period of time so as to realize the transfer to the $\ket{ggg}\ket{011}$ state. Such dynamics must occur only after half of the population initially stored in the $\ket{rrg}\ket{000}$ and $\ket{grr}\ket{000}$ states has been transferred to the $\ket{ggg}\ket{101}$ state. The remapping of the modes to different transitions is achieved by chirping the cavity modes or by shifting the atomic energies.

The correspondence of the cavity modes and the three-atomic transition frequencies, 
is shown in Fig. \ref{Diag}. We have devised a protocol to manage the entanglement process effectively, conceptualizing it as a two-stage procedure with time intervals $[t_0,t_1]$ and $[t_2,t_3]$, corresponding to stages 1 and 2 respectively. 
By the end of stage 1, when $t=t_1$, the following conditions must be satisfied, 
\begin{equation}
	\begin{split}
		&\rho_{rgr,000}(t_1)=0\\
		&\rho_{ggg,110}(t_1)=\dfrac{1}{3}\\
		&\rho_{rrg,000}(t_1)=\rho_{grr,000}(t_1)=\dfrac{1}{6}\\
		&\rho_{ggg,101}(t_1)=\dfrac{1}{3},\label{Obj1} 
	\end{split}
\end{equation}
with the population of all other states equal to zero. The condition for the  state of the atom-cavity system by the end of stage 2, when $t=t_3$, is,
\begin{equation}
	\begin{split}
		&\Psi(t_3)=\Psi_1.\label{Obj2} 
	\end{split}
\end{equation}

To measure the proximity of the state at the end of stage 2 with $\psi_1$, we introduce the fidelity $\mathcal{F}=\abs{\braket{\Psi_1}{\Psi(t_3)}}^2$. Our goal is to find a protocol that maximizes this fidelity satisfying the aforementioned conditions \eqref{Obj1} and \eqref{Obj2}. 

While a brute force approach for discovering an optimal control protocol for populating the target state is conceivable, it tends to lack robustness for controlled population transfer, due to sensitivity to parameter fluctuations \cite{Robustparam}. In addition, these type of approaches deal with increasingly complex dynamics as the system dimension is increased, which can be addressed by using techniques, such as adiabatic elimination, that result in reduced dimension system dynamics \cite{Reduceddyn}. Instead, we adopt a step-wise method that implies a series of population transfers at two-level avoided crossings. 
The rest of our paper is organized as follows. In Sec. \ref{Sec5Model}, we describe our theoretical model and a subsystem decomposition with effective Hamiltonians generating the transfer dynamics required for our two-stage protocols. In Sec. \ref{Sec5CFSTIRAPad}, we describe our adiabatic protocol that makes use of the STIRAP and FSTIRAP techniques. In Sec. \ref{Sec5CFSTIRAPnonad}, we describe our non-adiabatic protocol where continuous chirping is used to generate controlled non-adiabatic transition probabilities at avoided crossings. In Sec. \ref{Sec5Impl}, we outline two strategies for experimental realization of the coupling rate modulation and chirping. In Sec. \ref{Sec5Conc}, we deliver our conclusion and summarize the key points of our paper with some important insights.


\section{Theoretical model}\label{Sec5Model}
The field interaction Hamiltonian describing three atoms trapped in the optical lattice and placed in the three-mode cavity reads,
\begin{equation}\label{Hamil}
	\begin{split}
		&H(t)=\sum_{i}\omega_i(t)a_i^{\dag}a_i+\sum_{j_2>j_1}V_{R}(\vec{r}_{j_1},\vec{r}_{j_2})\\
		&+\sum_{i,j}\left(g_{ij}(\vec{r}_i,t)e^{i(\omega_{0,j}-\omega_i)t}\sigma_j^{+}a_i+\text{H.c.}\right),
	\end{split}    
\end{equation}
where $\omega_i(t)=\alpha_i(t-t_{\alpha,i})$ is the instantaneous frequency for the mode $i$ 
having initial frequency $\omega_{i}$, $\omega_{0,j}$ is the transition frequency of atom $j$, $g_{ij}(\vec{r}_i,t)$ is the coupling rate of atom $j$ with the cavity mode $i$, and $V_R(\vec{r}_{j_1},\vec{r}_{j_2})$ is the energy shift term due to Rydberg-Rydberg interactions between atoms $j_1$ and $j_2$. It reads, 
\begin{equation}
	\sum_{j_2>j_1}V_{R}(\vec{r}_{j_1},\vec{r}_{j_2})=\sum_{j_1}\left(V_1\sigma_{j_1}^{rr}\sigma_{j_1+1}^{rr}+V_2\sigma_{j_1}^{rr}\sigma_{j_1+2}^{rr}\right),
\end{equation}
where $V_1$ is the energy shift due to nearest-neighbour Rydberg interactions and $V_2$ is the energy shift due to next to nearest-neighbour Rydberg interactions. 


We work with a Hilbert space spanned by a basis of 17 atom-photon states. 
The basis consists of states with zero photons - the zero photon states (0P),
\begin{equation}
	V_{0P}=\{\ket{rrg},\ket{rgr},\ket{grr}\}\oplus\{\ket{000}\},
\end{equation}
the one-photon states (1P),
\begin{equation}
	V_{1P}=\{\ket{rgg},\ket{grg},\ket{ggr}\}\oplus\{\ket{100},\ket{010},\ket{001}\}, 
\end{equation}
and the two-photon states (2P),
\begin{equation}
	V_{2P}=\{\ket{ggg}\}\oplus\{\ket{110},\ket{101},\ket{011},\ket{200},\ket{020}\}.
\end{equation}



\begin{figure}[h!]
	\centering	\includegraphics[width=0.9\columnwidth]{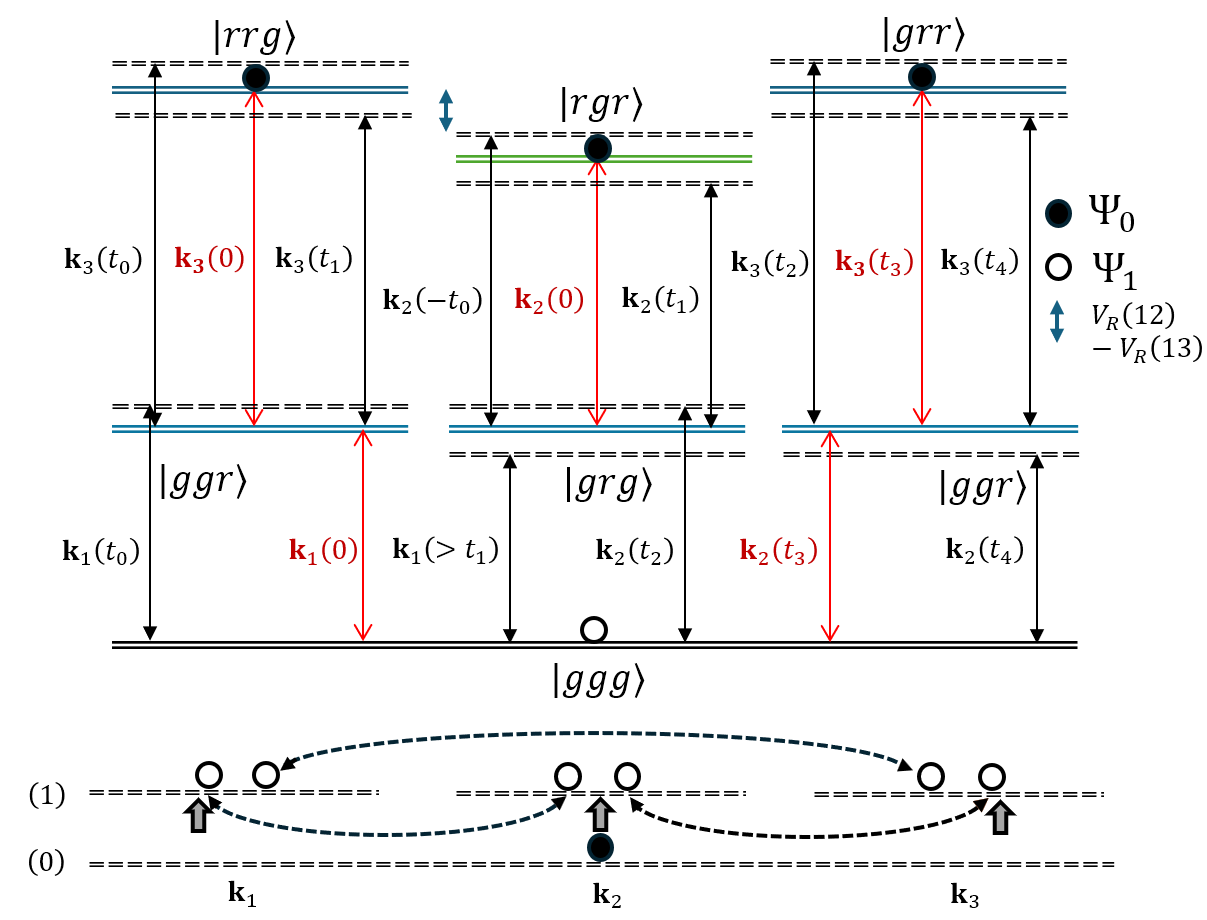}
	\caption{Atom-cavity setup showing the correspondence between cavity modes $i$ and transitions of each atom at different times as the mode frequencies $\omega_i=ck_i$ are modulated. The bottom of the graph shows Fock states of each cavity mode, treating each cavity mode as a qubit. The initial and final states $\Psi_0$ and $\Psi_1$ are shown in terms of the populations in each three-atom and cavity Fock state. The dashed curved arrows linking modes $\textbf{k}_i$ and $\textbf{k}_j$ correspond to the projection of $\Psi_1$ in the multi-mode state with a photon in each mode.
	}\label{Diag}
\end{figure}

We note that states $\ket{ggg,200},\ket{ggg,020}$ are  included in the basis due to a possibility of two-photon transitions within the 0P manifold, and one-photon transitions between populated states and the 1P states. The chirping of modes 1 and 2, as seen in Fig. \ref{Diag}, creates an interval in time where the effective two-photon transition rates between states in the 0P manifold, are sufficiently large to shift population within the manifold. 
We avoid populating these additional two-photon states by modulating the coupling rate. 
Besides, we assume a large one-photon detuning to prevent population of the 1P state manifold. 
Given this, we employ the method of adiabatic elimination \cite{Walls2008} to exclude the intermediate 1P states. 
In order to obtain an understanding of all the physical interactions present in the atom-cavity system, we introduce an effective two-photon Hamiltonian,
\begin{equation}
   H^{\text{eff}}(t)=\sum_{i}\omega_i(t)a_i^{\dag}a_i+H_S+H_T,\label{Ham333}
\end{equation}
where $H_S$ represents the dispersive contributions, including AC Stark shifts, and $H_T$ represents the near resonant two-photon processes, involving either absorption or emission into multiple modes, where,
\begin{align}\label{HamilStark}
\begin{split}
    &H_S=\\
    &a_1^{\dag}a_1\left(\dfrac{\abs{g_1}^2}{\Delta_1}A_{1,+}+\dfrac{\abs{g_1}^2}{\Delta_1-V_2}A_{2,+}+\dfrac{\abs{g_1}^2}{\Delta_1-V_1}A_{3,+}\right)\\
    &-a_1a_1^{\dag}\left(\dfrac{\abs{g_1}^2}{\Delta_1}A_{1,-}+\dfrac{\abs{g_1}^2}{\Delta_1-V_2}A_{2,-}+\dfrac{\abs{g_1}^2}{\Delta_1-V_1}A_{3,-}\right)\\
    &+a_2^{\dag}a_2\left(\dfrac{\abs{g_2}^2}{V_2+\Delta_2}A_{1,+}+\dfrac{\abs{g_2}^2}{\Delta_2}A_{2,+}+\dfrac{\abs{g_2}^2}{\Delta_2+V_2-V_3}A_{3,+}\right)\\
    &-a_2a_2^{\dag}\left(\dfrac{\abs{g_2}^2}{V_2+\Delta_2}A_{1,-}+\dfrac{\abs{g_2}^2}{\Delta_2}A_{2,-}+\dfrac{\abs{g_2}^2}{\Delta_2+V_2-V_1}A_{3,-}\right)\\
    &+a_3^{\dag}a_3\left(\dfrac{\abs{g_3}^2}{V_1+\Delta_2}A_{1,+}+\dfrac{\abs{g_3}^2}{\Delta_2+V_3-V_2}A_{2,+}+\dfrac{\abs{g_3}^2}{\Delta_3}A_{3,+}\right)\\
    &-a_3a_3^{\dag}\left(\dfrac{\abs{g_3}^2}{V_1+\Delta_2}A_{1,-}+\dfrac{\abs{g_3}^2}{\Delta_2+V_3-V_2}A_{2,-}+\dfrac{\abs{g_3}^2}{\Delta_3}A_{3,-}\right),
\end{split}
\end{align}

\begin{align}\label{HamilStarkcoeffs}
\begin{split}
    &A_{1,+}=3\sigma_1^{11}\sigma_2^{11}\sigma_3^{11},\\
    &A_{1,-}=\sum_{i=1,2,3}\sigma_i^{+}\sigma_1^{11}\sigma_2^{11}\sigma_3^{11}\sigma_i^{-},\\
    &A_{2,+}=\sum_{i=1,3}\sigma_i^{+}\sigma_1^{11}\sigma_2^{11}\sigma_3^{11}\sigma_i^{-},\\
    &A_{2,-}=2\sigma_1^{22}\sigma_2^{11}\sigma_3^{22},\\
    &A_{3,+}=(\sigma_2^{+}\cdot\sigma_2^{-}+\sum_{i=1,2,3}\sigma_i^{+}\cdot\sigma_i^{-})\sigma_1^{11}\sigma_2^{11}\sigma_3^{11},\\
    &A_{3,-}=\sigma_2^{22}\left(2(\sigma_1^{22}\sigma_3^{11}+\sigma_3^{22}\sigma_1^{11})+\sigma_1^{+}\sigma_3^{-}+\sigma_3^{+}\sigma_1^{-}\right),
\end{split}
\end{align}

\begin{align}\label{HamilTwoPhoton}
\begin{split}
    &H_T=\dfrac{g^{*}_2g^{*}_1}{\Delta_2}e^{i(\Delta_2+\Delta_1)t}H_{T,12}+\dfrac{g^{*}_3g^{*}_1}{\Delta_3}e^{i(\Delta_3+\Delta_1)t}H_{T,13},\\
    &+\dfrac{g^{*}_3g^{*}_2}{\Delta_3}e^{i(\Delta_3+\Delta_2)t}H_{T,23}+\text{H.c.},\\
    &H_{T,12}=2a_1^{\dag}a_2^{\dag}\sigma_{1}^{-}\sigma_2^{11}\sigma_{3}^{-},\\
    &H_{T,13}=a_1^{\dag}a_3^{\dag}\sigma_2^{-}\left(\sigma_{1}^{-}\sigma_{3}^{11}+\sigma_{1}^{11}\sigma_{3}^{-}\right),\\
    &H_{T,23}=a_2^{\dag}a_3^{\dag}\sigma_2^{-}\left(\sigma_{1}^{-}\sigma_{3}^{11}+\sigma_{1}^{11}\sigma_{3}^{-}\right).
\end{split}
\end{align}

The operator formalism is instrumental in describing how the atom-cavity interactions, depicted in Fig. \ref{Diag}, 
not only generate two-photon transitions between pairs of  states in the $V_{0P}$ and $V_{2P}$ manifolds, but also induce dynamical Stark shifts. 
These Stark shifts could contribute to a detrimental time-varying deviation from two-photon resonance \cite{STIRAPVitreview}, resulting  in a more complex evolution of the adiabatic energies and mixing angles compared to conventional three-level STIRAP. Therefore the STIRAP and FSTIRAP transfers must explicitly account for these energy shifts. We also note that $A_{3,-}$ has terms $\sigma_1^{+}\sigma_3^{-}+\sigma_3^{+}\sigma_1^{-}$ that shift  population within the 0P manifold. It causes a splitting between superpositions of states $\ket{rrg}$ and $\ket{grr}$, such as 
$\dfrac{\ket{rrg}\pm\ket{grr}}{\sqrt{2}}$, breaking the degeneracy as a result of an effective coupling within the 0P manifold. 
Furthermore, Hamiltonian (\ref{Ham333}) reveals that not all states in the $0P$ and $2P$ manifolds interact with each other. This observation justifies partitioning the atom-cavity system into decoupled subsystems. 

The entanglement transfer is performed in two stages by means of a series of decoupled processes, 
each occurring in specific subsystems involving distinct subspaces of the atom-cavity state space. 
We define the subsystem $\mathcal{H}_{i,j}$ corresponding to stage $i$ and transfer process $j$, as the class $\{\{v_{i,j}\},H^{\text{eff}}_{i,j}\}$, where $\{v_{i,j}\}$ is the state basis that undergoes dynamics 
generated by the effective Hamiltonian $H^{\text{eff}}_{i,j}$. $H^{\text{eff}}_{i,j}$ is obtained by projecting the effective Hamiltonian $H^{\text{eff}}(t)$ \eqref{Ham333} onto the state basis $\{v_{i,j}\}$ after applying a unitary transformation. The effective Hamiltonians are described in terms of the two-photon effective coupling rates, $g_{i,kl}^{\text{eff}}$, and the two-photon effective detunings, $\Delta^{\text{eff}}_{i,kl}$ (where $k,l$ denote the modes involved in the couplings/detunings), see Appendix \ref{Sec5Adiab} for more details. 
We obtain the following stage-wise subsystem decomposition for the total system,

\begin{eqnarray*}
\begin{split}
    &\mathcal{H}=\mathcal{H}_{1,1}\oplus\mathcal{H}_{1,2}\oplus\mathcal{H}_{1}^{\prime} && t\in[t_0,t_1]\\
    &\mathcal{H}=\mathcal{H}_{2,1}\oplus\mathcal{H}_{2,2}\oplus\mathcal{H}_{2}^{\prime} && t\in[t_2,t_3], 
\end{split}
\end{eqnarray*}
where 
$\mathcal{H}_{1,1}$ has state basis $\{\ket{rgr,000}, \ket{ggg,110}\}$,   $\mathcal{H}_{1,2}$ has state basis $\{\ket{rrg,000}, \ket{grr,000}, \ket{ggg,101}\}$,  $\mathcal{H}_{2,1}$ has state basis $\{\ket{ggg,101}, \ket{ggg,110}\}$, and $\mathcal{H}_{2,2}$ has state basis $\{\ket{rrg,000}, \ket{grr,000}, \ket{ggg,011}\}$.
The subsystems $\mathcal{H}_{i}^{\prime}$  consist of states outside those belonging to the $\mathcal{H}_{i,j}$'s.  These states are invariant under the action of $H^{\text{eff}}(t)$ \eqref{Ham333} for the corresponding time interval. 

We find that for the subsystems retaining more than two levels after adiabatic elimination, one of the states, what we call a dark state, can be decoupled from the other two by using the Morris-Shore transformations \cite{Zlatanov2020}. This is a result of the previously mentioned degeneracy breaking of states $\ket{rrg,000}$ and $\ket{grr,000}$ by the effective coupling in $H_S$ \eqref{HamilStark}. The transformations take us to a state basis $\{\ket{\tilde{0}},\ket{\tilde{1}},\ket{\tilde{2}}\}$ where $\ket{\tilde{0}}$ is the uncoupled dark state. This implies that all processes during the entanglement transfer can ultimately be described using only two-level systems (TLSs), where the wavefunctions of the three-state subsystems, $\mathcal{H}_{i,2}$, only populate the two coupled states $\ket{\tilde{1}}, \ket{\tilde{2}}$. 
For simplicity, we substitute the Morris-Shore transformed Hamiltonians, $H_{i,2}^{\text{eff,MS}}(t)$, for the three-level effective Hamiltonians $H_{i,2}^{\text{eff}}(t)$, see Appendix \ref{Sec5Adiabdiab} for details. 
%
Furthermore, we use the formalism of diabatic and adiabatic states to describe the subsystems' dynamics in the vicinity of 
the two-level system avoided crossings. In the next section, we describe the adiabatic protocol for entanglement transfer. We used the QuTiP package in Python to carry out numerical analysis \cite{QuTiP}.

\section{Adiabatic Protocol}\label{Sec5CFSTIRAPad}
The defining feature of adiabatic evolution of a system is that the system remains in the same eigenstate of the Hamiltonian throughout the time evolution. Depending on the path taken in parameter space, the eigenstate can end up in a single basis state or a coherent superposition of states. The relevant parameter of interest for STIRAP/FSTIRAP with the super-effective two-level system, spanned by the basis $\{\ket{a},\ket{b}\}$, is the mixing angle $\varphi$- which gives us information about the non-adiabatic coupling, $\dot{\varphi}$, and the evolution within each eigenstate (adiabatic state) $\ket{w_{\pm}(t)}=\sin\left(\varphi(t)-(1\mp 1)\pi/2\right)\ket{a}+\cos\left(\varphi(t)-(1\mp 1)\pi/2\right)\ket{b}$. We choose parameters that satisfy the adiabaticity condition $\abs{\dot{\varphi}}\ll \abs{2E_{\text{ad}}}$, where $E_{\text{ad}}$ is the adiabatic energy, while also satisfying the STIRAP and FSTIRAP mixing angle conditions, given below,

\begin{eqnarray}
\begin{split}
    &\lim_{t\rightarrow+\infty}\varphi(t)=\pi/2 && \text{(STIRAP)}\\
    &\lim_{t\rightarrow+\infty}\varphi(t)=\pi/4  && \text{(FSTIRAP)}
\end{split}\label{mixconds}
\end{eqnarray}

In this protocol, we adiabatically transfer populations between the 0P and 2P states for each subsystem $\mathcal{H}_{i,j}$. For stage 1, we carry out a STIRAP process for $\mathcal{H}_{1,1}$ and a FSTIRAP process for $\mathcal{H}_{1,2}$. The STIRAP process results in a complete population transfer from $\ket{rgr,000}$ to $\ket{ggg,110}$, and the FSTIRAP process results in half population transfer from $\ket{rrg,000}$ and $\ket{grr,000}$ to $\ket{ggg,101}$, as can be inferred from the above mixing angle conditions \eqref{mixconds}. We modulate the coupling rates $g_{i}(t)$ in stage 1 so that the processes occur in separate time intervals, where the coupling rate $g_{1}(t)$ is separated into two terms, $g_{1,s1}(t)$ and $g_{1,s2}(t)$, each of which are set to zero when the other has non-trivial magnitude. We have separated the processes in time to ensure that the dispersive coupling doesn’t introduce unwanted Stark shifts that interfere with each process. The term $g_{1,s1}(t)$ acts as the Stokes pulse for the STIRAP transfer and $g_{2}(t)$ acts as the pump pulse. The term $g_{1,s2}(t)$ acts as a sequence of two Stokes pulses for the FSTIRAP transfer and $g_{3}(t)$ act as the pump pulse. 

For stage 2, the evolution for both subsystems $\mathcal{H}_{2,i}$ will be adiabatic. For $\mathcal{H}_{2,1}$, the 2P states, $\ket{ggg,101}$ and $\ket{ggg,110}$, acquire dynamical phases adiabatically. For $\mathcal{H}_{2,2}$, the remaining population in the $\ket{rrg,000}$ and $\ket{grr,000}$ states are transferred to $\ket{ggg,011}$ through a STIRAP process with the Stokes pulse generated by $g_2(t)$ and the pump pulse generated by $g_3(t)$. Next, we describe the adiabatic protocol in detail.

The system is initialized in the Rydberg W state \eqref{Psii} at $t_0=0$. The stage 1 coupling rates and mode frequencies are, 

\begin{equation}
	\begin{split}
		&g_1(t)=g_{1,s1}(t)+g_{1,s2}(t)\\
        &\equiv A_{s1}e^{-(t^{\prime}-2.5\tau_{s}-t_{s1})^2/(2\tau_{s}^2)}\\
        &+A_{s2}(e^{-(t^{\prime}+t_{s2})^2/(2\tau_{s}^2)}+e^{-(t^{\prime}-t_{s2})^2/(2\tau_s^2)}),\\
		&g_2(t)=A_{p1}e^{-(t^{\prime}-2.5\tau_{s}-t_{p1})^2/(2\tau_{s}^2)},\\
		&g_3(t)=A_{p2}e^{-(t^{\prime}-t_{s2})^2/(2\tau_{s}^2)},\\
		&\omega^{\prime}_1(t)=-\alpha_{0}(t^{\prime}-t_{\alpha}),\\
		&\omega^{\prime}_3(t)=\omega^{\prime}_2(t)=-\omega^{\prime}_1(t). 
	\end{split}\label{FST11}
\end{equation}
where $t^{\prime}=t-5\tau_s$ is the offset time coordinate chosen such that all stage 1 dynamics occurs in the interval $[t_0,t_0+20\tau_s]$. The STIRAP process occurs in the interval $[t_0+10\tau_s,t_0+20\tau_s]$ and the FSTIRAP process occurs in the interval $[t_0,t_0+10\tau_s]$. The ordering of the two transfer processes in time, as given here, is not a strict requirement and can be reversed. The times $t_i$ are time-offsets for the Gaussian functions in the coupling rates and linear chirp functions in the mode frequencies. The Gaussian functions have FWHM $\sqrt{2}\tau_{i}$ and peak amplitudes $A_{i}$. The linear chirp functions have chirp rates $\alpha_{i}$. 

\begin{figure}[h!]
	\centering
	\subfigure[]{\includegraphics[width=0.9\columnwidth]{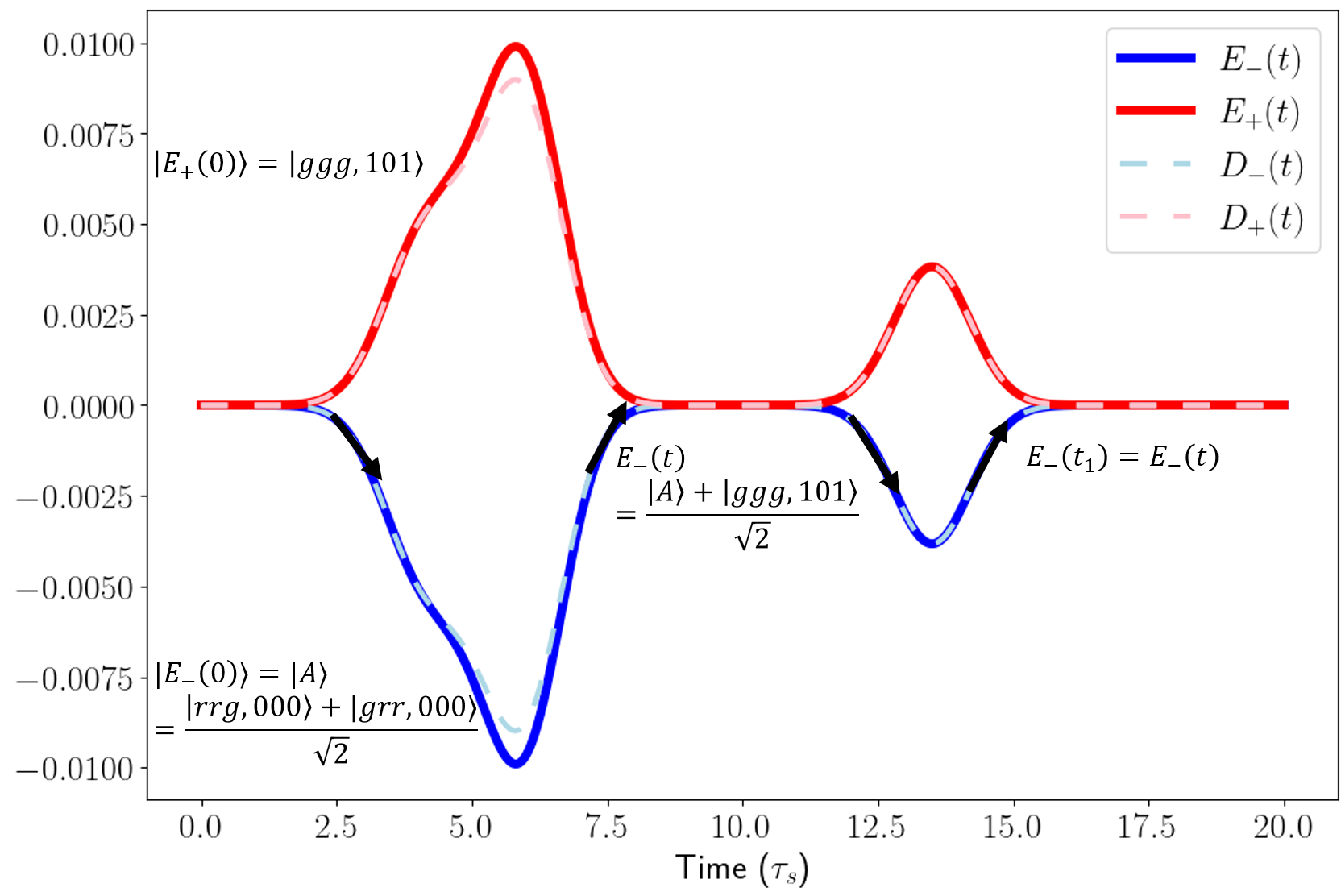}}
	
	\subfigure[]{\includegraphics[width=0.9\columnwidth]{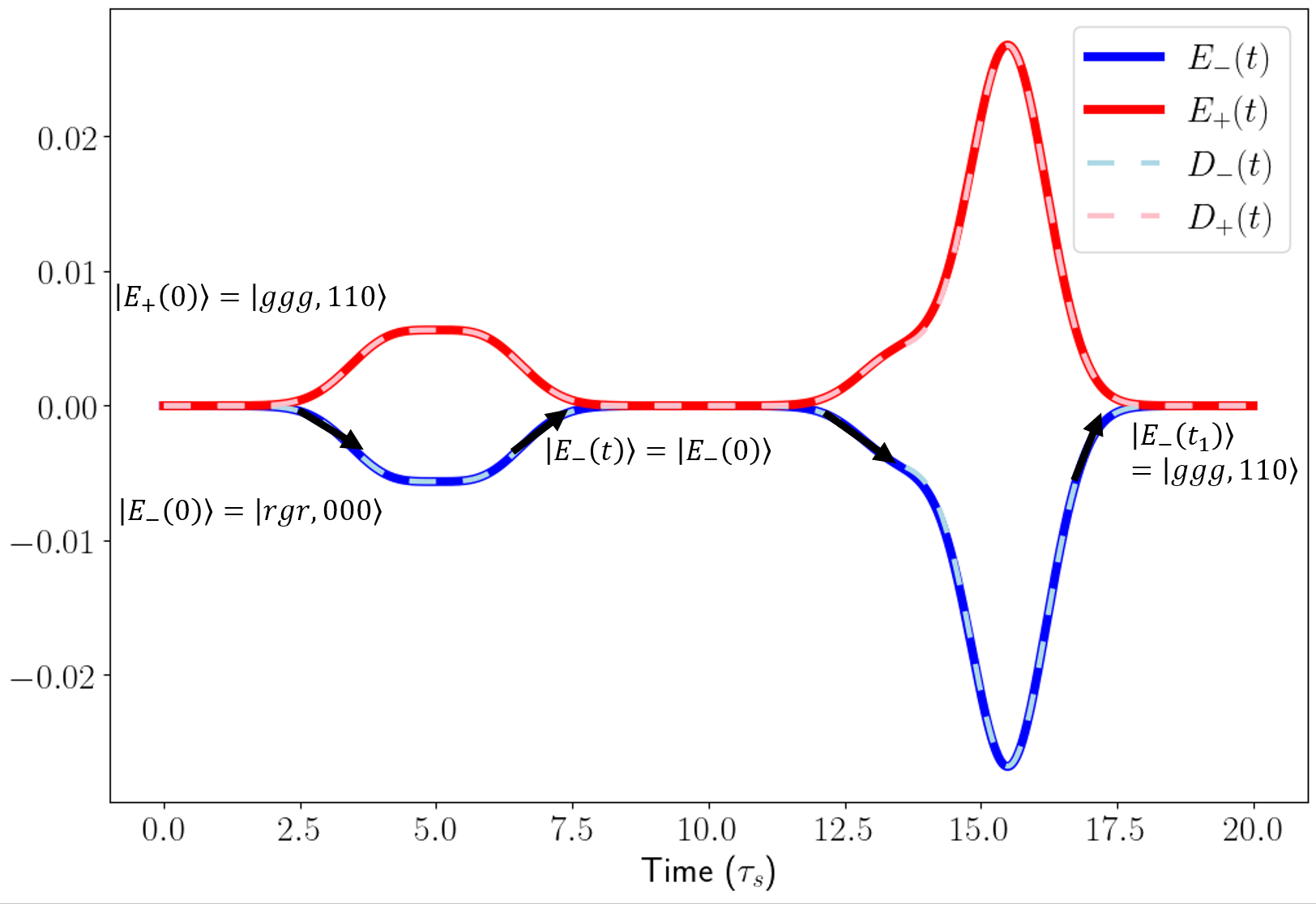}}
	\caption{Plots of the diabatic ($D_{\pm}(t)$) and adiabatic state ($E_{\pm}(t)$) energies for the supereffective TLS corresponding to $\mathcal{H}_{1,1}$ (a) and $\mathcal{H}_{1,2}$ (b) for the adiabatic protocol. 
    Black arrows denote the system trajectory. Stage 1 parameters: $A_{s1}=A_{s2}=A_{p2}=0.505$, $A_{p1}=1.634$, $\tau_{s}=1000$, $t_{s1}=t_{s2}=\tau_{s}$, $t_{p1}=3t_s$, $\alpha_0=0$, $-\Delta_1=\Delta_3=\Delta_2=100$, $V_1=2V_2=10\Delta_2$.}\label{fig: Stage1adiabener}
\end{figure}

\begin{figure}[ht!]
	\centering	\includegraphics[width=0.9\columnwidth]{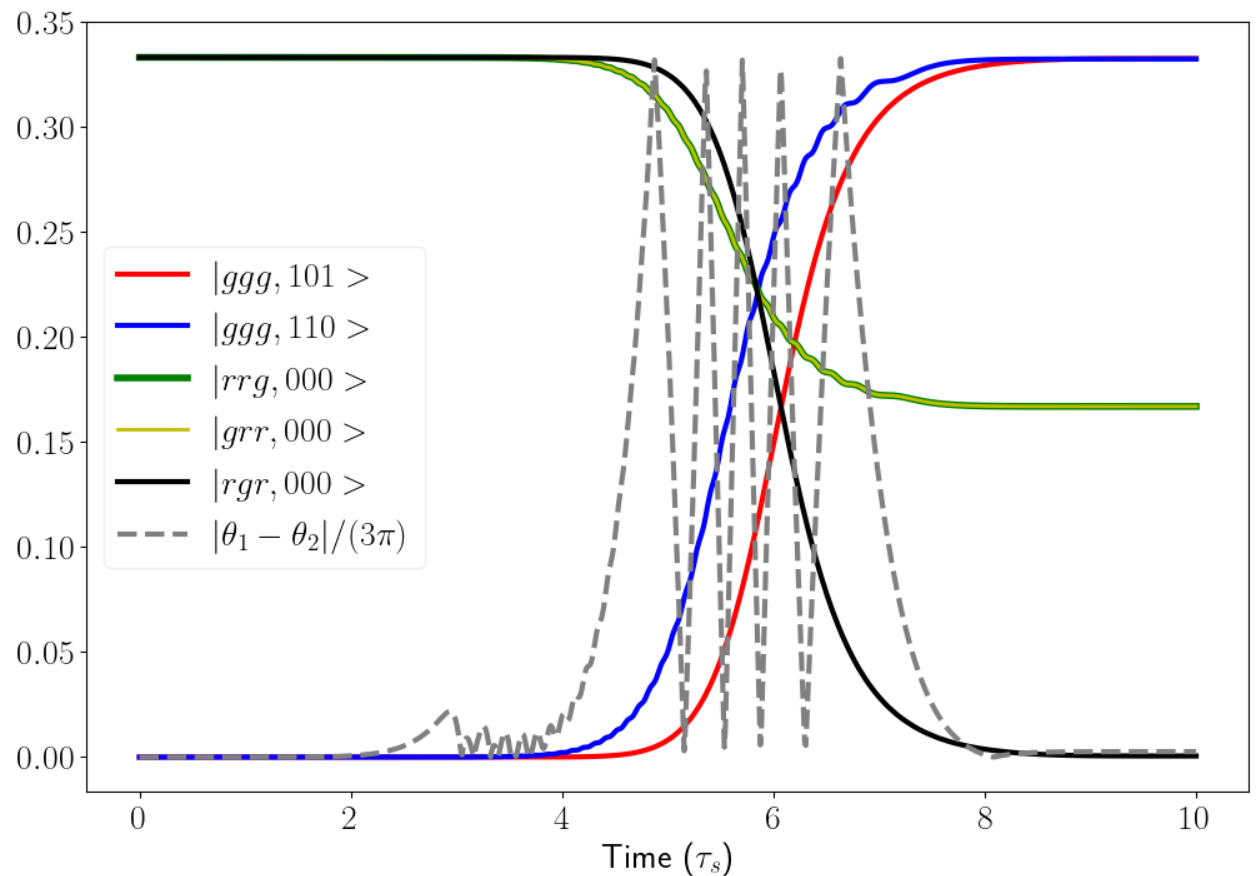}
	\caption{Evolution of the populations and relative phase for stage 1 of the adiabatic protocol. Parameters are the same as in for Fig. \ref{fig: Stage1adiabener}} \label{fig: Stage1Adiab}
\end{figure}

With our couplings and mode frequencies known, we used the expressions for the effective couplings and detunings in the effective Hamiltonians $H^{\text{eff}}_{i,j}$ to calculate the diabatic (the diagonal terms of the super-effective TLS Hamiltonian) and adiabatic energies. 
The evolution of the diabatic and adiabatic energies for both subsystems is given in Fig. \ref{fig: Stage1adiabener}, demonstrating the STIRAP/FSTIRAP protocols in action with the 
time evolution of the populations corresponding to the states in $\mathcal{H}_{1,1}$ and $\mathcal{H}_{1,2}$ shown in Fig. \ref{fig: Stage1Adiab}. As our choice of parameters, given in the caption of Fig. \ref{fig: Stage1adiabener}, satisfies the mixing angle conditions \eqref{mixconds}, we achieve the stage 1 population transfer objectives in \eqref{Obj1} with completely adiabatic dynamics. However, the population transfer is not the only aspect we are concerned with, as we also have to control the local phases in the wavefunction at the end of stage 1 (at $t=t_1=20\tau_s$). For describing the adiabatic evolution of the phases, we use the interaction picture wavefunction at $t=t_1$ given by,
\begin{align}
\begin{split}
    &\psi_I(t_1;\{\theta_i\})=\dfrac{1}{\sqrt{3}}\left(e^{i\theta_1}\ket{ggg,110}\right.\\
    &\left.+e^{i\theta_2}\dfrac{\ket{rrg,000}+\ket{grr,000}}{2}+e^{i\theta_3}\ket{ggg,101}\right).   
\end{split}\label{Target1}
\end{align}
Since the evolution is adiabatic, the local phases $\theta_i$ depend on the eigenenergies of their corresponding effective Hamiltonians, see Appendix \ref{Sec5Adiabdiab} for details. Our choice of parameters must satisfy the constraint $\theta_1\equiv\theta_2\text{mod}(2\pi)$ where,
\begin{align}
\begin{split}
    &\theta_1=\int_{t_0}^{t_1}dt^{\prime}\text{ }\sqrt{\left(\Delta_{1,12}^{\text{eff}}(t^{\prime})\right)^2+\left(g_{1,12}^{\text{eff}}(t^{\prime})\right)^2}+\Delta^0_{1,12}(t^{\prime})\\
    &-\left(\omega_1(t^{\prime})+\omega_2(t^{\prime})\right),\\
    &\theta_2=\int_{t_0}^{t_1}dt^{\prime}\text{ }\sqrt{\left(\Delta_{1,13}^{\text{eff}}(t^{\prime})\right)^2+2\left(g_{1,13}^{\text{eff}}(t^{\prime})\right)^2}\\
    &+\Delta^0_{1,13}(t^{\prime})+g_{1,33}^{\text{eff}}(t^{\prime})/2,\\
    &\theta_3=\theta_2-\int_{t_0}^{t_1}dt^{\prime}\text{ }\left(\omega_1(t^{\prime})+\omega_3(t^{\prime})\right).
\end{split}\label{Phase1}
\end{align}

As seen in Fig. \ref{fig: Stage1Adiab}, the evolution of the relative phase $\abs{\theta_1-\theta_2}$ demonstrates our success in achieving the aforementioned phase constraint. We've achieved a fidelity of $0.9989$ with respect to the target state $\psi(t_1;\{0\})$. During the time interval $[t_1,t_2]$, we chirp modes 1 and 2 to prepare for stage 2, chirping the mode frequencies negatively so that mode 2 will be in resonance with the three-atom transition $\ket{ggr}\rightarrow\ket{ggg}$, and mode 1 will be sufficiently far off detuned that it will no longer interact with the atomic system. The coupling rates and mode frequencies for stage 2 are, 

\begin{equation}
	\begin{split}
		&g_1(t)=0,\\
		&g_2(t)=A_{s^{\prime}}e^{-(t^{\prime\prime}+t_{s^{\prime}})^2/(2\tau_{s^{\prime}}^2)},\\
		&g_3(t)=A_{p^{\prime}}e^{-(t^{\prime\prime}-t_{s^{\prime}})^2/(2\tau_{s^{\prime}}^2)},\\
		&\omega^{\prime}_1(t)=-D,\\
		&\omega^{\prime}_3(t)=-\alpha_{0^{\prime}}(t^{\prime\prime}-t_{\alpha,2}),\\
		&\omega^{\prime}_2(t)=-\omega^{\prime}_3(t)-(V_{12}+2\Delta_2).
	\end{split}\label{FST21}
\end{equation}

\begin{figure}[h!]
	\centering	\includegraphics[width=0.9\columnwidth]{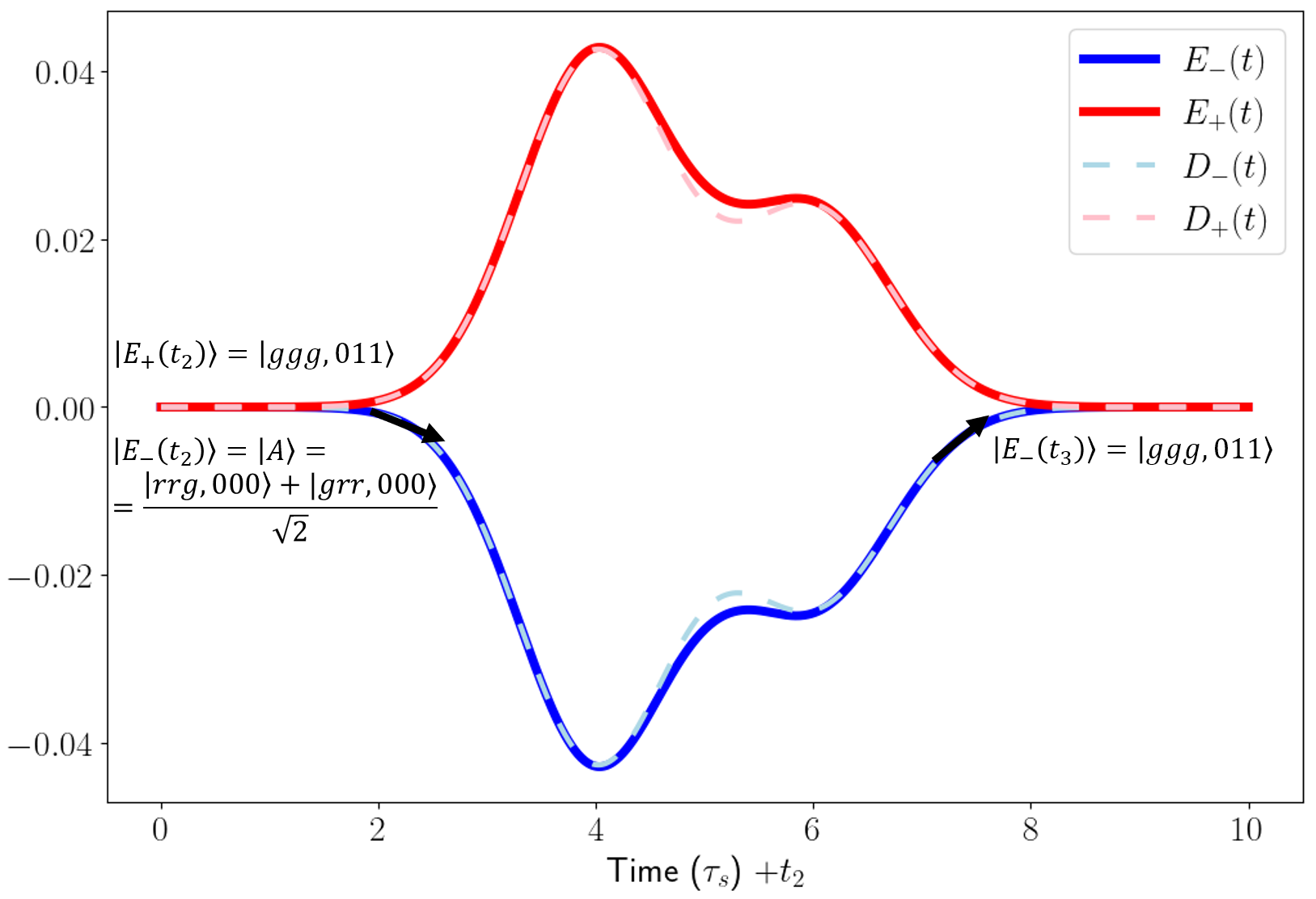}
	\caption{Plots of the adiabatic state energies for the supereffective TLS corresponding to $\mathcal{H}_{2,2}$ for the adiabatic protocol. STIRAP is used for complete population transfer. Black arrows denote the system trajectory. Parameters: $A_{s^{\prime}}=1.677$, $A_{p^{\prime}}=1.25$, $t_{s^{\prime}}=\tau_{s^{\prime}}=1000$, $\alpha_{0^{\prime}}=0$, $-\Delta_2=\Delta_3=100$, $V_1=2V_2=10\Delta_2$.} \label{fig: STIRAPadiabPh2}
\end{figure}

\begin{figure}[h!]
	\centering
	\subfigure[]{\includegraphics[width=0.9\columnwidth]{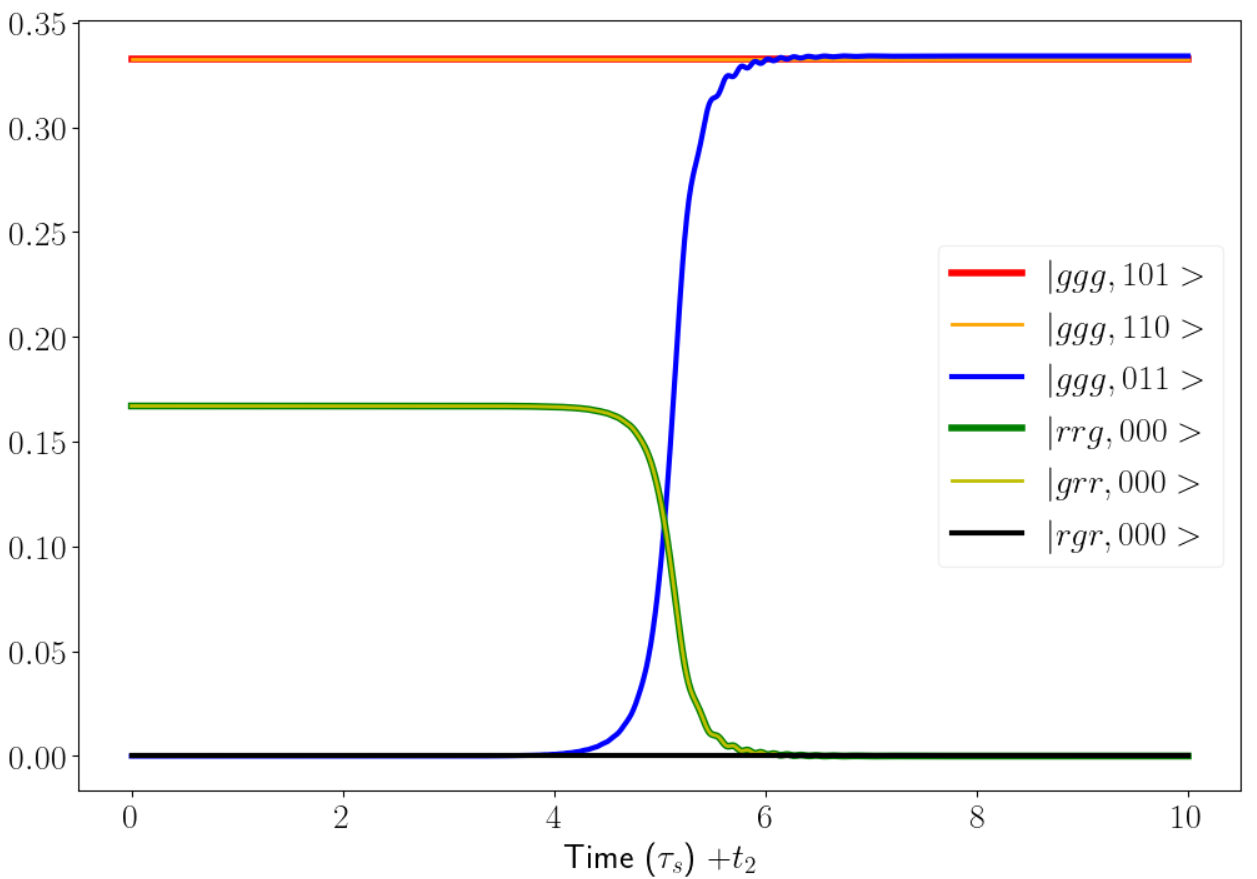}}
	
	\subfigure[]{\includegraphics[width=0.9\columnwidth]{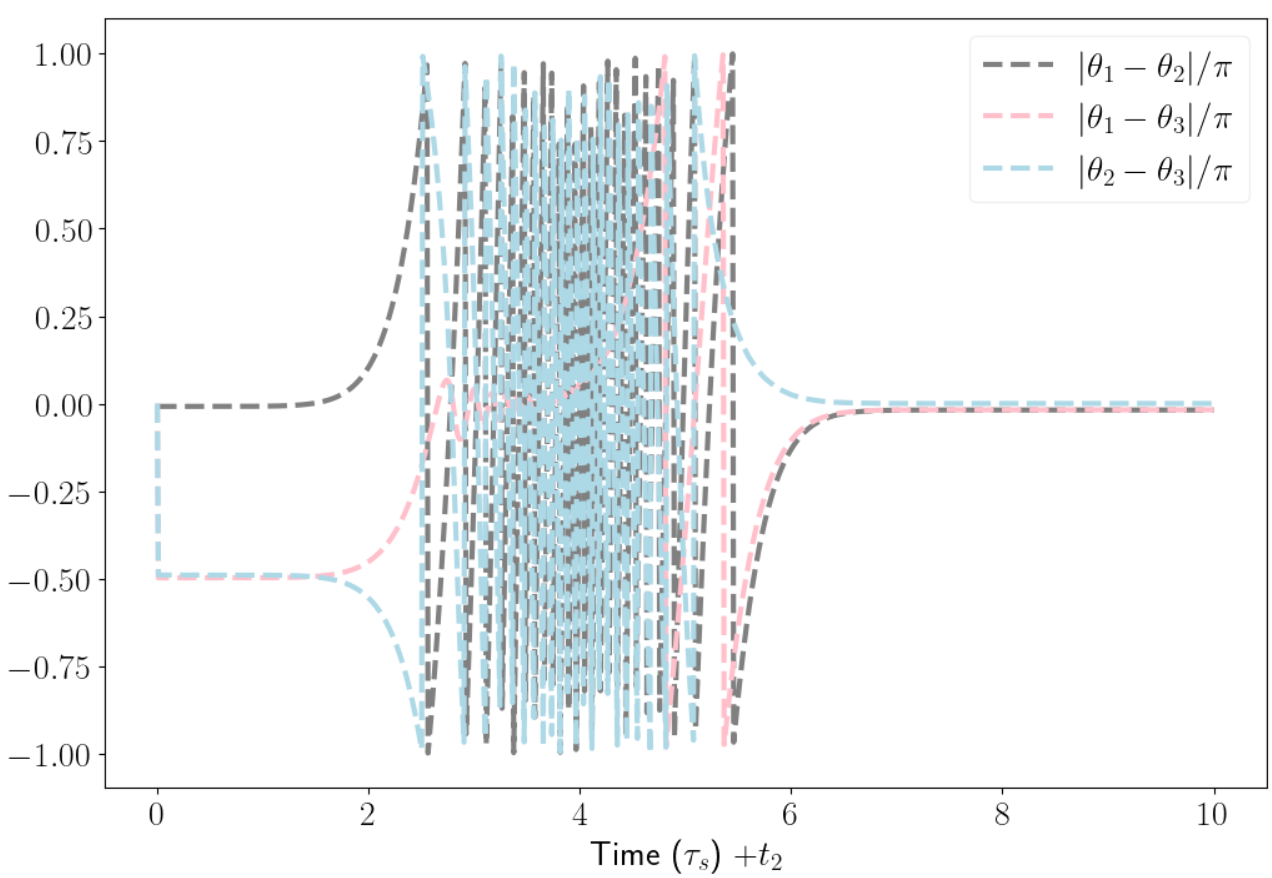}}
	\caption{Evolution of the populations (a) and relative phases (b) for stage 2 of the adiabatic protocol. Parameters are the same as in Fig. \ref{fig: STIRAPadiabPh2}}\label{fig: Phase2evol}
\end{figure}

Where $t^{\prime\prime}=t-(t_2+5\tau_{s^{\prime}})$ is the offset time coordinate chosen such that all stage 2 dynamics occurs in the interval $[t_2,t_2+10\tau_{s^{\prime}}]$. During stage 2, while all evolution is adiabatic, see Fig. \ref{fig: STIRAPadiabPh2}, the components of the wavefunction in the 2P states develop different phases. The stage 2 time evolution of the populations and phases, corresponding to the states in subsystems $\mathcal{H}_{2,1}$ and $\mathcal{H}_{2,2}$, is shown in Fig. \ref{fig: Phase2evol}. The final state is given by,

\begin{align}
\begin{split}
    &\psi(t_3;\{\theta_i\}, \{\phi_i\})=\dfrac{1}{\sqrt{3}}\left(e^{i(\theta_1+\phi_1)}\ket{ggg,110}\right.,\\
    &\left.+e^{i(\theta_2+\phi_2)}\ket{ggg,101}+e^{i(\theta_3+\phi_3)}\ket{ggg,011}\right).    
\end{split}\label{Target2}
\end{align}

As in the case of stage 1, the local phases $\phi_i$ depend on the eigenenergies of the effective Hamiltonians. To maximize the fidelity, we must select parameters that satisfy the constraint $(\theta_1+\phi_1)\equiv(\theta_2+\phi_2)\text{mod}(2\pi)\equiv(\theta_2+\phi_3)\text{mod}(2\pi)$, where,
\begin{align}
\begin{split}
    &\phi_1=-\int_{t_2}^{t_3}dt^{\prime}\text{ }\dfrac{3\abs{g_2(t^{\prime})}^2}{\Delta_2},\\
    &\phi_2=-\int_{t_2}^{t_3}dt^{\prime}\text{ }\dfrac{3\abs{g_3(t^{\prime})}^2}{\Delta_2},\\
    &\phi_3=\int_{t_2}^{t_3}dt^{\prime}\text{ }\sqrt{\left(\Delta_{2,23}^{\text{eff}}(t^{\prime})\right)^2+2\left(g_{2,23}^{\text{eff}}(t^{\prime})\right)^2}\\
    &+\Delta^0_{2,23}(t^{\prime})+g_{2,33}^{\text{eff}}(t^{\prime})/2-\left(\omega_2(t^{\prime})+\omega_3(t^{\prime})\right).
\end{split}\label{Phases2}
\end{align}

With our choice of parameters, given in the caption of Fig. \ref{fig: Phase2evol}, we successfully achieve our objective for stage 2, achieving a fidelity of $0.9991$ with the target state $\Psi_1=\psi(t_3;\{0\},\{0\})$. The protocol we have developed here shows that entanglement transfer from the three-atom Rydberg W-state to the three-mode photonic W-state is achieved with purely adiabatic evolution. We note that the STIRAP technique is very sensitive to the Stark shifts generated by the dispersive coupling, requiring us to use independent sequences for the STIRAP and FSTIRAP processes in stage 1. The introduction of counterdiabatic terms \cite{STAcount} in the Hamiltonian would address this concern and greatly improve the robustness of the adiabatic protocol. This limitation is not so profound in our non-adiabatic protocol, which we introduce in the next section.   

\section{Non-adiabatic protocol}\label{Sec5CFSTIRAPnonad}
Compared to adiabatic evolution, non-adiabatic transitions require that the Hamiltonian of a system has non-trivial couplings between the eigenstates for some period of time. In cases like the Landau-Zener (LZ) system, we can ensure a series of non-adiabatic transition occur at fixed times with defined non-adiabatic transition rates, each occurring at an engineered avoided crossing. The non-adiabatic transition probabilities depend on the mixing angle of the super-effective TLS through the coupling between adiabatic states, $\dot{\varphi}$. 

In this protocol, we continuously chirp the system, by using non-zero chirp rates $\alpha_i$ to generate a sequence of two-level avoided crossings between the TLS eigenstates, $\ket{w_{\pm}(t)}$, for each of the effective subsystems $\mathcal{H}_{i,j}$. 
By using chirp rates that do not cancel out in the effective detunings, $\Delta^{\text{eff}}_{i,kl}(t)$, and using coupling rates, $g_{i}(t)$, that generate 2P effective coupling rates, $g^{\text{eff}}_{i,kl}(t)$, that peak at zeroes of  $\Delta^{\text{eff}}_{i,kl}(t)$, we sweep through the 2P resonances- generating LZ-like avoided crossings.  The evolution for subsystems $\mathcal{H}_{1,1}$ and $\mathcal{H}_{2,2}$ will be adiabatic to ensure complete population transfer. 
During stage 1, for $\mathcal{H}_{1,1}$, we obtain complete adiabatic population transfer from $\ket{rgr,000}$ to $\ket{ggg,110}$ by choosing parameters that increase the adiabatic energy gap and completely suppress non-adiabatic transitions at the avoided crossing. For $\mathcal{H}_{1,2}$, we must determine suitable parameters that lead to a non-adiabatic transition probability of half so that we end up in an equal superposition of the initial states, $\ket{rrg,000}$ and $\ket{grr,000}$, and final state, $\ket{ggg,101}$. In contrast to the adiabatic protocol, we designed the spectrum of $H^{\text{eff}}_{1,2}$ in the non-adiabatic protocol such that the eigenstates terminate exclusively in either the initial or final basis states rather than an equal superposition of both initial and final states, hence our requirement to evolve our wavefunction to populate two eigenstates/adiabatic states. 
For stage 2, the evolution for both subsystems $\mathcal{H}_{2,i}$ will be adiabatic where the 2P states, $\ket{ggg,101}$ and $\ket{ggg,110}$, in $\mathcal{H}_{2,1}$ acquire dynamical phases adiabatically, and the remaining population in the $\ket{rrg,000}$ $\ket{grr,000}$ states will be transferred to $\ket{ggg,011}$ at an avoided crossing where the non-adiabatic transition probability is zero.

The same initial configuration is used as that of the adiabatic protocol in Sec. \ref{Sec5CFSTIRAPad} and the time intervals for each stage and time coordinates $t^{\prime}$, $t^{\prime\prime}$ are also defined analogously. However, the time interval for stage 1 is shorter, with pertinent dynamics occurring for $t\in[t_0,t_1]$ and $t_1=t_0+10\tau_s$. Next, we proceed to describe the non-adiabatic protocol in detail. 

The stage 1 coupling rates and mode frequencies are,
\begin{equation}
	\begin{split}
		&g_1(t)=A_{s}e^{-(t^{\prime}+t_{s})^2/(2\tau_{s}^2)},\\
		&g_2(t)=A_{p1}e^{-(t^{\prime}-t_{p1})^2/(2\tau_{p1}^2)},\\
		&g_3(t)=A_{p2}e^{-(t^{\prime}-t_{p2})^2/(2\tau_{p2}^2)},\\
		&\omega^{\prime}_1(t)=-\alpha_{0,1}(t^{\prime}-t_{\alpha}),\\
		&\omega^{\prime}_2(t)=-\alpha_{0,2}(t^{\prime}-t_{\alpha}),\\
		&\omega^{\prime}_3(t)=-\alpha_{0,3}(t^{\prime}-t_{\alpha}).
	\end{split}\label{FST}
\end{equation}

\begin{figure}[h!]
	\centering
	\subfigure[]{\includegraphics[width=0.9\columnwidth]{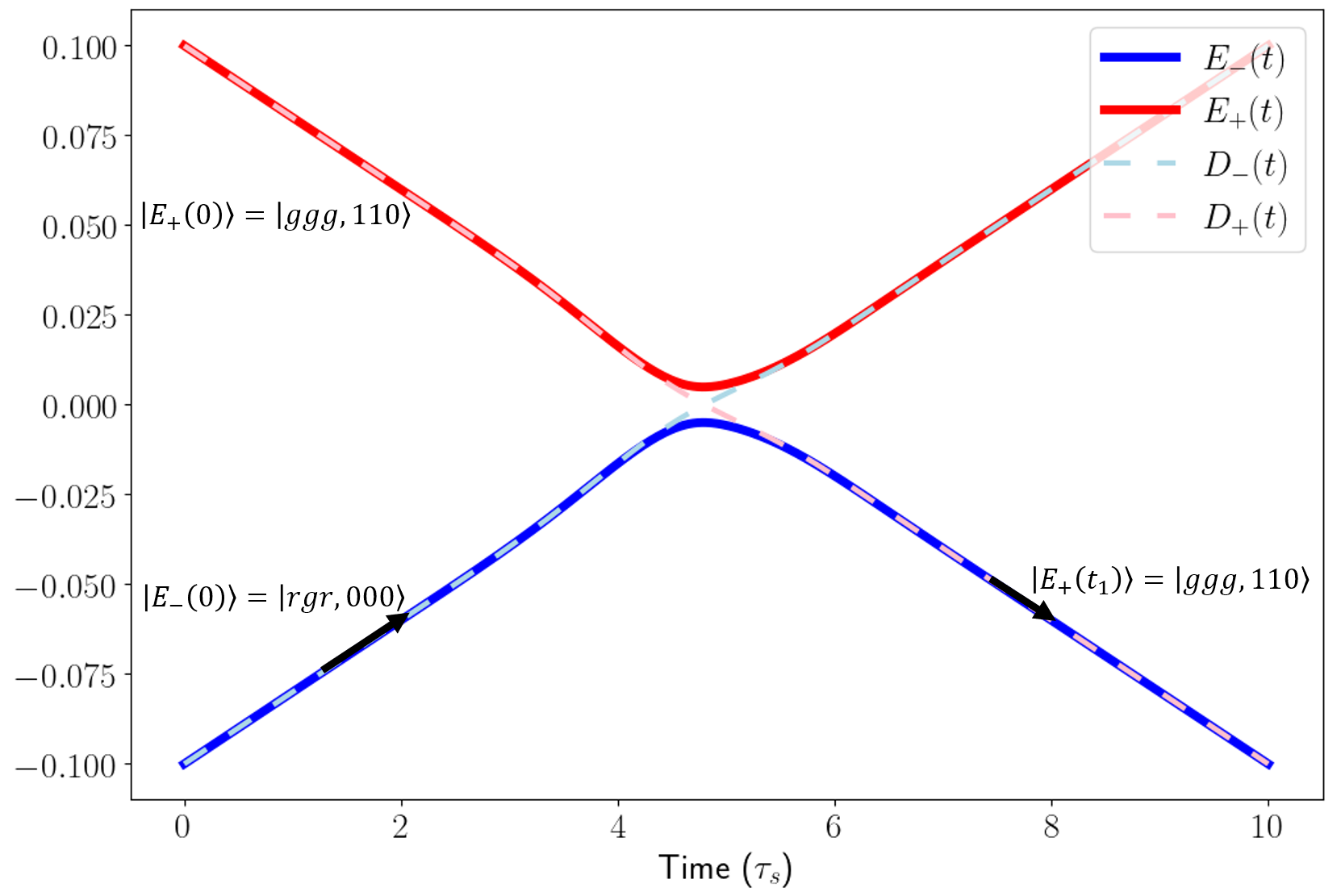}}
	
	\subfigure[]{\includegraphics[width=0.9\columnwidth]{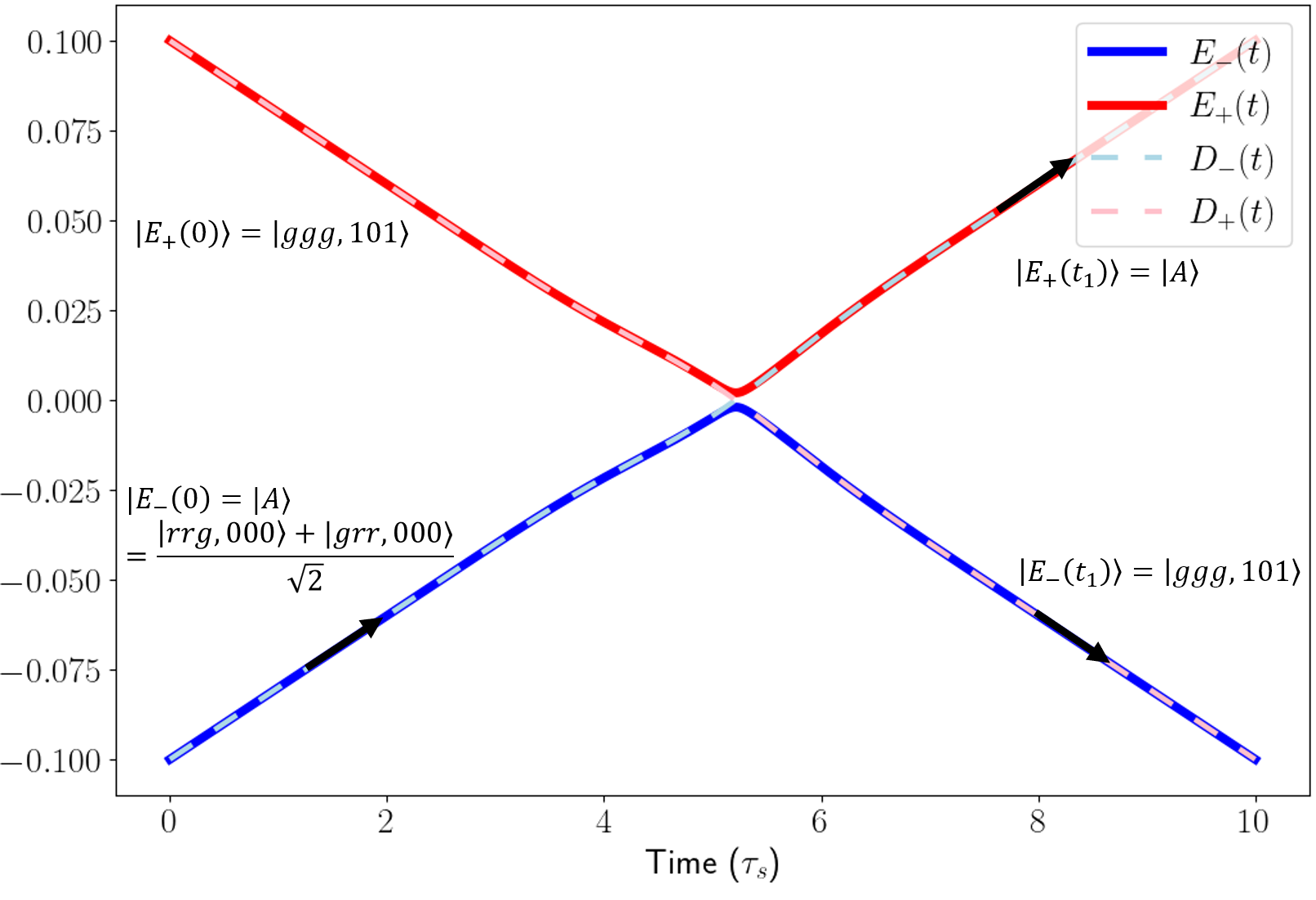}}
	\caption{Plots of the diabatic ($D_{\pm}$) and adiabatic ($E_{\pm}$) state energies for the supereffective TLS corresponding to $\mathcal{H}_{1,1}$ (a) and $\mathcal{H}_{1,2}$ (b) for the non-adiabatic protocol. 
    Black arrows denote the system trajectory. Parameters: $A_s=0.55$, $A_{p1}=0.925$, $A_{p2}=0.285$, $\tau_{p1}=\tau_{p2}=\tau_{s}=1000$, $t_{s}=t_{p2}=t_{\alpha}=0$, $t_{p1}=-336.7$, $\alpha_{0,i}=2\text{E}-5$, $\Delta_1=100$, $V_2=2V_2=10\Delta_1$.}\label{fig: phase1evo1}
\end{figure}

\begin{figure}[h!]
	\centering
	\subfigure[]{\includegraphics[width=0.9\columnwidth]{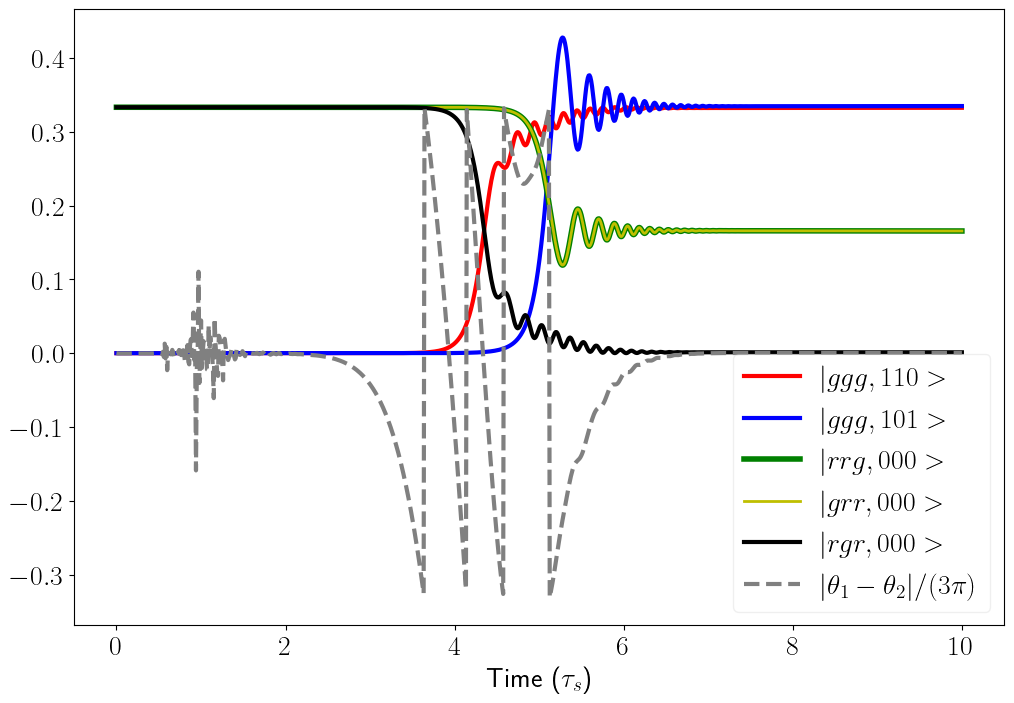}}

	\caption{Evolution of the populations and relative phase for stage 1 of the non-adiabatic protocol. Parameters are the same as in Fig. \ref{fig: phase1evo1}.}\label{fig: phase1evo2}
\end{figure}

For $\mathcal{H}_{1,1}$, non-adiabatic transitions are suppressed by increasing the energy gap between eigenstates through choosing a large amplitude for $g_2(t)$. 
In addition, we impose a constraint similar to the pulse converging constraint in FSTIRAP to ensure that the Stark shifts due to modes 1 and 2 cancel out uniformly as we move away from the avoided crossing,
\begin{equation}
	t_{p2}=\sqrt{\abs{\left(\dfrac{t_s\tau_s}{\tau_{p2}}\right)^2-2\tau_{p2}^2\log\abs{\dfrac{A_{p2}}{A_{s}}}}}.\label{tp2cond}
\end{equation}
Fig. \ref{fig: phase1evo1} shows the adiabatic energy and population evolution of $\mathcal{H}_{1,1}$ for the parameter set given in the figure caption. 
For $\mathcal{H}_{1,2}$, the aforementioned parameters results in a small enough adiabatic gap and steep enough slope at the avoided crossing that we successfully generate the non-adiabatic transition required for equal population of the two adiabatic states, as the time evolution of the populations in Fig. \ref{fig: phase1evo2} shows. An approximate expression for the non-adiabatic transition probability, in the case where the contributions from the Stark shifts are much smaller than those from the chirp at the avoided crossing, is given by $P=\exp\left(-2\pi\dfrac{\left(\dfrac{2A_{p3}A_s}{\Delta_1}\right)^2}{\alpha_0}\right)$, see Appendix \ref{App5A} for the derivation.
For the evolution of the local phases, the wavefunction and 
phase $\theta_1$ will have the same form as seen in Eqs. \eqref{Target1} and \eqref{Phase1} respectively. However, the phases $\theta_{2/3}$ are defined differently due to the presence of the non-adiabatic transition, which will introduce a non-Abelian phase $\theta_{S}$ (the Stokes phase) that doesn't have a general expression for time-dependent Rabi frequencies. The phases are,
\begin{align}
\begin{split}
    &\theta_2=\theta_{\text{ad}}-\theta_{S},\\   
    &\theta_3=-\theta_{\text{ad}}-\int_{t_0}^{t_1}dt^{\prime}\text{ }\left(\omega_1(t^{\prime})+\omega_3(t^{\prime})\right),\\
    &\theta_{\text{ad}}=\left(\int_{t_0}^{t_-}+\int_{t_+}^{t_1}\right)dt^{\prime}\text{ }\sqrt{\left(\Delta_{1,13}^{\text{eff}}(t^{\prime})\right)^2+2\left(g_{1,13}^{\text{eff}}(t^{\prime})\right)^2}\\
    &+\Delta^0_{1,13}(t^{\prime})+g_{1,33}^{\text{eff}}(t^{\prime})/2.
\end{split}\label{Phase1nonadiab}
\end{align}
Where $t_{\pm}$ refer to the times that bound the interval for which $\theta_{S}$ is calculated, outside of which the evolution is adiabatic. 
We achieved a fidelity of 0.9988 with the target state $\psi(t_1;0,0)$. 

\begin{figure}[ht]
	\centering
    \includegraphics[width=0.9\columnwidth]{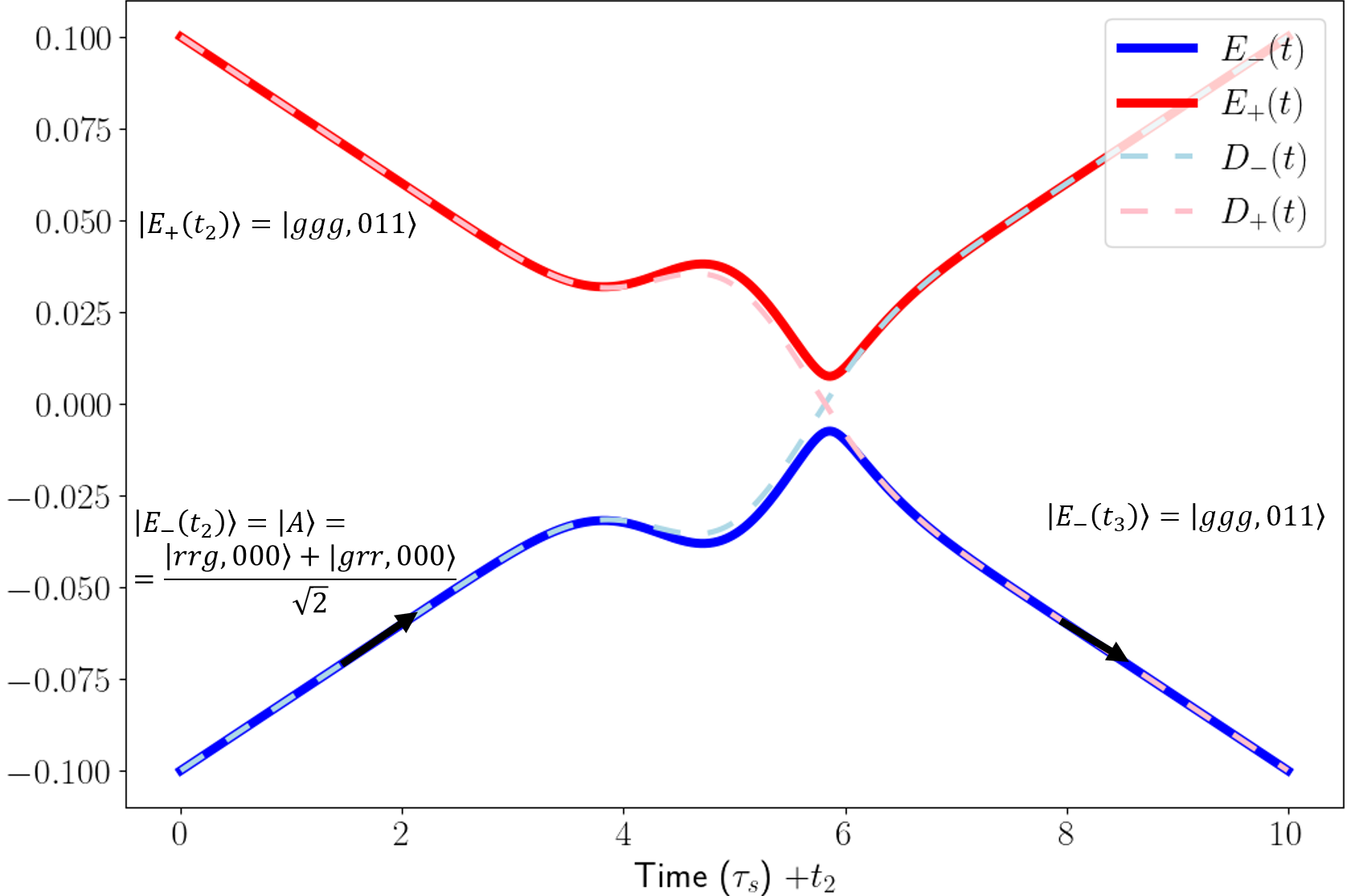}
	\caption{Plots of the diabatic ($D_{\pm}$) and adiabatic ($E_{\pm}$) state energies for the supereffective TLS corresponding to $\mathcal{H}_{2,2}$ for the non-adiabatic protocol. Black arrows denote the system trajectory. Parameters: $A_{s^{\prime}}=0.983$, $A_{p3^{\prime}}=1.03$, $\tau_{p^{\prime}}=\tau_{s^{\prime}}=1000$, $t_{s^{\prime}}=t_{\alpha}=0$, $t_{p^{\prime}}=2$, $\alpha^{\prime}_{0,i}=2\text{E}-5$, $\Delta_1=100$, $V_2=2V_2=10\Delta_1$.}\label{fig: phase2evo2}
\end{figure}

\begin{figure}[ht]
	\centering
	\subfigure[]{\includegraphics[width=0.9\columnwidth]{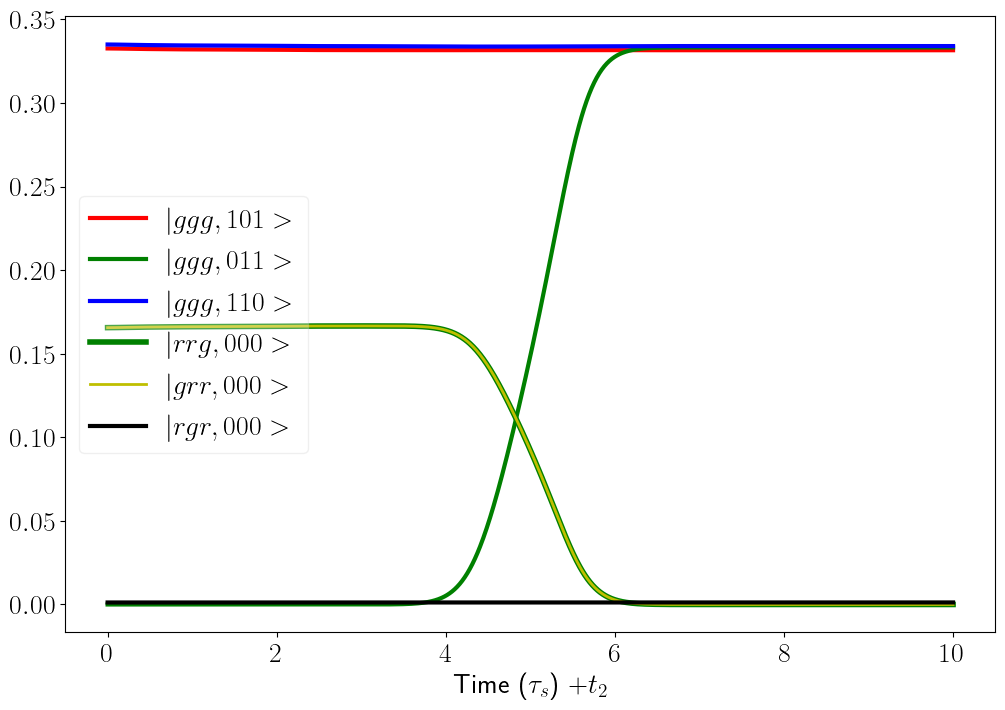}}
    \subfigure[]{\includegraphics[width=0.9\columnwidth]{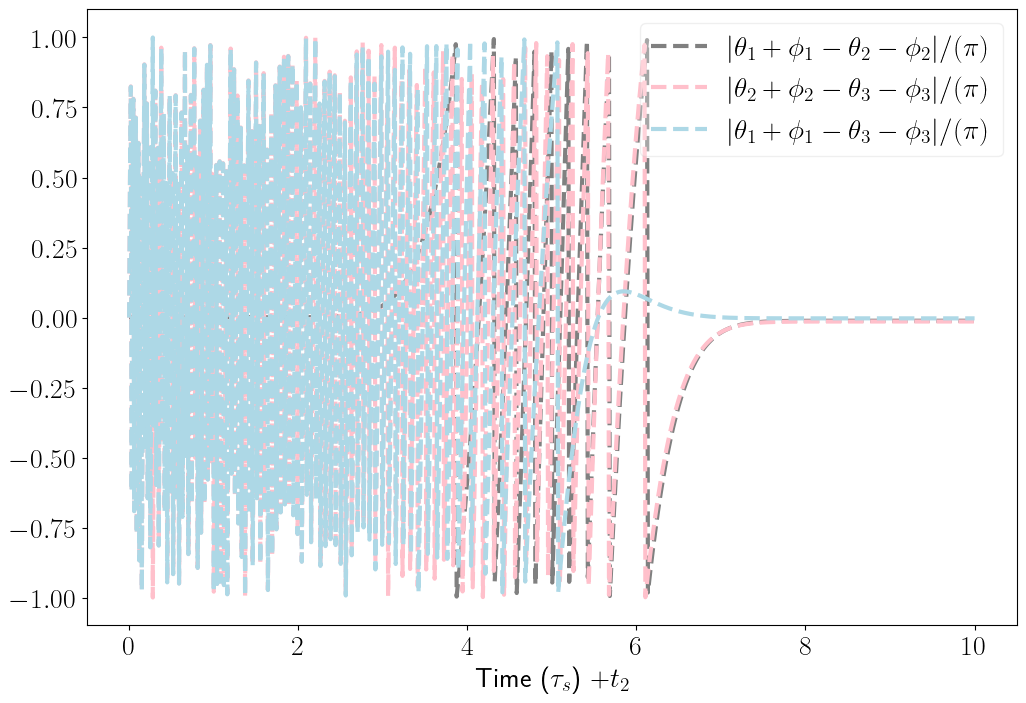}}
	\caption{Evolution of the populations (a) and relative phases (b) for stage 2 of the non-adiabatic protocol. Parameters are the same as in Fig. \ref{fig: phase2evo2}.}\label{fig: phase2pops}
\end{figure}

Stage 1 is terminated at $t_1=10\tau_s$. 
During the time interval $[t_1,t_2]$, we chirp modes 1 and 2 to prepare for stage 2, chirping the mode frequencies negatively so that mode 2 will be resonant with the three-atom transition $\ket{ggr}\rightarrow\ket{ggg}$, and mode 1 will be sufficiently far off detuned that it will no longer interact with the atomic system. For stage 2, the coupling rates and mode frequencies are,
\begin{equation}
	\begin{split}
		&g_1(t)=0,\\
		&g_2(t)=A_{s^{\prime}}e^{-(t^{\prime\prime}+t_{s^{\prime}})^2/(2\tau_{s^{\prime}}^2)},\\
		&g_3(t)=A_{p^{\prime}}e^{-(t^{\prime\prime}-t_{p^{\prime}})^2/(2\tau_{p^{\prime}}^2)},\\
		&\omega^{\prime}_1(t)=-D,\\
		&\omega^{\prime}_2(t)=-\alpha_{0,2}^{\prime}(t^{\prime\prime}-t_{\alpha,2}),\\
		&\omega^{\prime}_3(t)=\alpha_{0,3}^{\prime}(t^{\prime\prime}-t_{\alpha,2}).
	\end{split}\label{FST2}
\end{equation}
The process of stage 2 proceeds in a similar way to that of the adiabatic protocol where all evolution is completely adiabatic. For $\mathcal{H}_{2,1}$, the probability amplitudes of the two 2P states $\ket{ggg,101}$ and $\ket{ggg,110}$ acquire dynamical phases adiabatically. $\mathcal{H}_{2,2}$ has complete adiabatic population transfer to the 2P state $\ket{ggg,011}$, as seen in Fig. \ref{fig: phase2pops}. The target objective and phases for stage 2 of the non-adiabatic protocol are the same as given by Eqs. \eqref{Target2} and \eqref{Phases2} except that we have to use Eq. \eqref{Phase1nonadiab} for $\theta_{2/3}$. With our choice of parameters, given in the caption for Fig. \ref{fig: phase2evo2}, we succeeded in setting all relative phases to zero, and achieving a fidelity of 0.9984 with the target state $\Psi_1$. 


Although we have shown the constraint conditions required to obtain zero relative phase for each component of the photonic W-state, it may be the case that we want to correct the phases after the entanglement transfer. We may use local phase gates to act on each cavity mode $k$ through operation $U\rho(\psi(t_3))U^{\dag}=\rho(\psi_1)$ where $U=\exp\left(i\left(\sum_i\Theta_ia_i^{\dag}a_i\right)\right)$, $\rho$ is the density matrix operator, and the phases are given by,
\begin{align}
\begin{split}
    &\Theta_1=2n_1\pi-(\theta_1+\phi_1),\\
    &\Theta_2=2n_2\pi-(\theta_2+\phi_2),\\
    &\Theta_3=2n_3\pi-(\theta_3+\phi_3).
\end{split}
\end{align}
Where $n_i\in\mathbb{Z}$. In the assumption that our protocol has a noiseless implementation and the populations in $\rho(\psi(t_3))$ are the same as those in $\rho(\Psi_1)$, the fidelity of the protocol is then given by the fidelity of the phase gate.

Both protocols successfully achieves the target populations and local phases for the 2P cavity modes required by objective $\eqref{Psif}$. We find that the non-adiabatic protocol proves as effective in generating the photonic W-state as the adiabatic protocol with a reduced process time and higher tolerance of dispersive coupling facilitated by chirping the cavity modes during the population transfer. There is still a question of determining an expression for the Stokes phase $\theta_S$. Although we can easily numerically calculate the individual phases for the photonic W-state, an analytic result, similar to our derived expression for the non-adiabatic transition probability, would be very useful. 
In the next section, we discuss two strategies to implement our protocols in the lab.

\section{Strategy for Experimental Realization}\label{Sec5Impl}
\begin{figure}[h!]
	\centering
	\includegraphics[width=0.9\columnwidth]{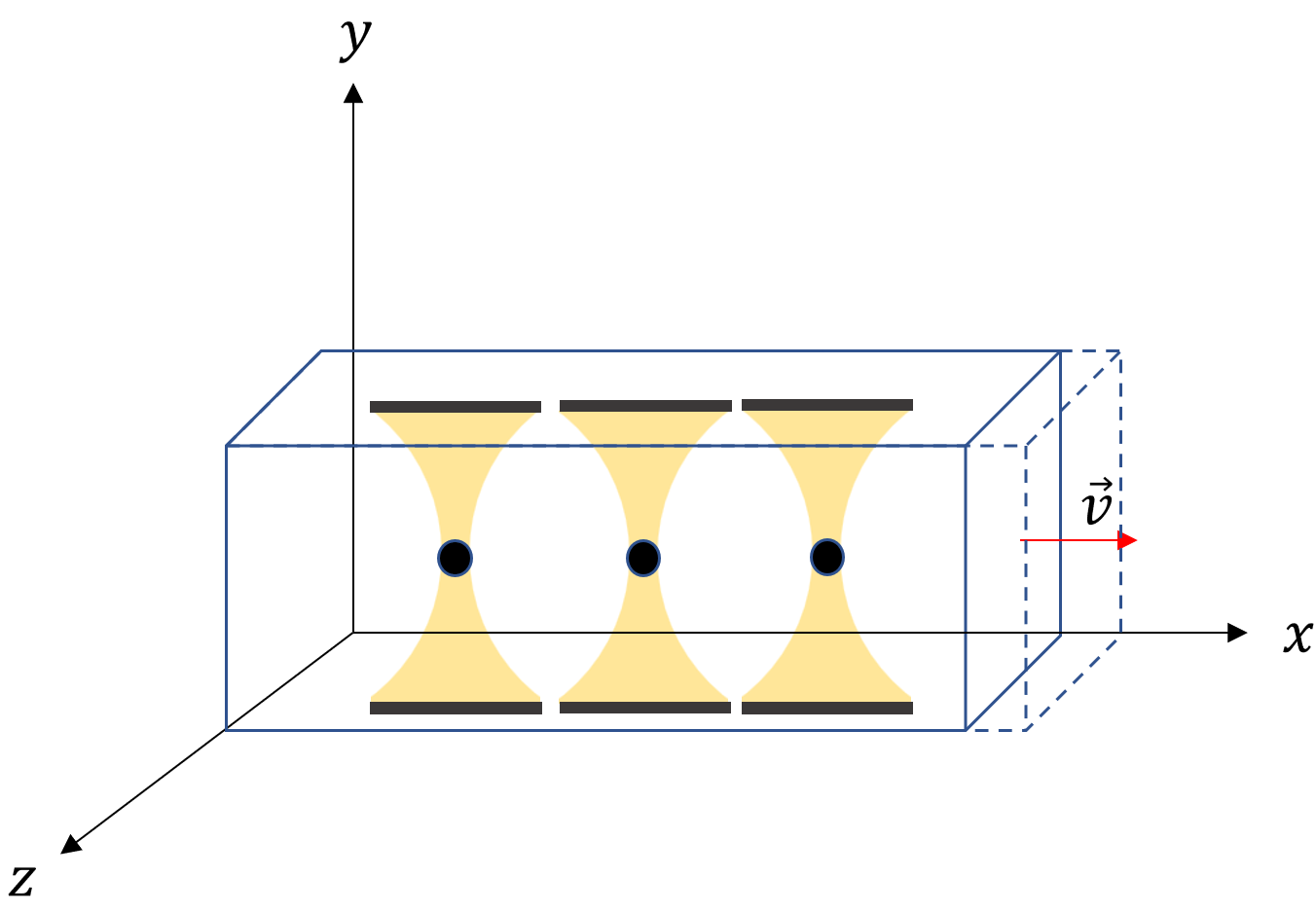}
	\caption{The 3D rectangular cavity with a moving mirror along one coordinate axis. The three atoms are trapped using optical tweezers to guide the atoms between nodes and anti-nodes of each of the 3 cavity modes to induce the Gaussian modulation of the mode coupling, $g_i(t)$.}\label{fig: 3Dcav}
\end{figure}

We have developed two protocols that would be of interests for experimental realization. Both protocols requires chirping of the cavity modes at some point in time as well as modulation of the cavity coupling rates, both of which are not trivial tasks to achieve. We introduce a method to implement the coupling modulations by using a 3D rectangular cavity with a single moving mirror- with motion along an optical axis of the cavity, see Fig. \ref{fig: 3Dcav}. The three atoms, initially existing in a Rydberg W-state, are trapped inside the cavity using optical tweezers- realizing a dynamically programmable array of atoms \cite{Bluvstein2022} where each atom can be transported between nodes and anti-nodes of cavity modes. We therefore generate coupling rates, $g_j(t)$, that follow the spatial trajectories of the atoms. We consider the limiting case where the speed $\abs{v}$ of the moving mirror is non-relativistic and where the maximum absolute change in the length of an optical axis is much smaller than the initial length ($\abs{vT}\ll L_0$). The axis of motion is the $x$-axis. In this limit, we obtain the below results for the cavity mode frequency, $\omega_{\vec{k}}(t)$, and positive frequency modes, $\phi_{\vec{k}}^{(\mu)}(t)$,

\begin{equation}
	\begin{split}
		&\omega_{\vec{k}}(t)\approx\sqrt{k_y^2+k_z^2+k_x(0)^2}\left(1-t\dfrac{vk_x(0)^2}{k_y^2+k_z^2+k_x(0)^2}\right)\\
		&\phi_{\vec{k}}^{(\mu)}(\vec{r},t)=\sqrt{\dfrac{8}{L_x(t)L_yL_z}}\sin(k_x(t)x)\sin(k_yy)\sin(k_zz)\\
  &\times\dfrac{1}{\sqrt{2\omega_{\vec{k}}(t)}}e^{-i\omega_{\vec{k}}t}
	\end{split}
\end{equation}

A derivation of this result is present in Appendix \ref{App5B}. Through a suitable choice of the cavity dimensions, $(L_x(t),L_y,L_z)$, the mirror velocity $\vec{v}$, and assignment of the mode numbers, $(n_x,n_y,n_z)$, to each mode $j$, we can fix the initial frequencies and chirp rates required to successfully complete the two-stage entanglement transfer. Chirping the cavity modes will coincide with a period of expanding the cavity $x$-dimension from $L_x(t_i)$ to $L_x(t_f)$ for the interval $[t_i,t_f]$. Shuttling the atoms between nodes and anti-nodes of the cavity modes using optical tweezers will give us the required coupling rates for our protocols. The use of programmable atomic arrays in this setup also provides an additional means of controlling the chirp rates by the addition of electric and magnetic fields to shift the atomic levels' energies. 

\begin{figure}[h!]
	\centering
	\subfigure[]{\includegraphics[width=0.9\columnwidth]{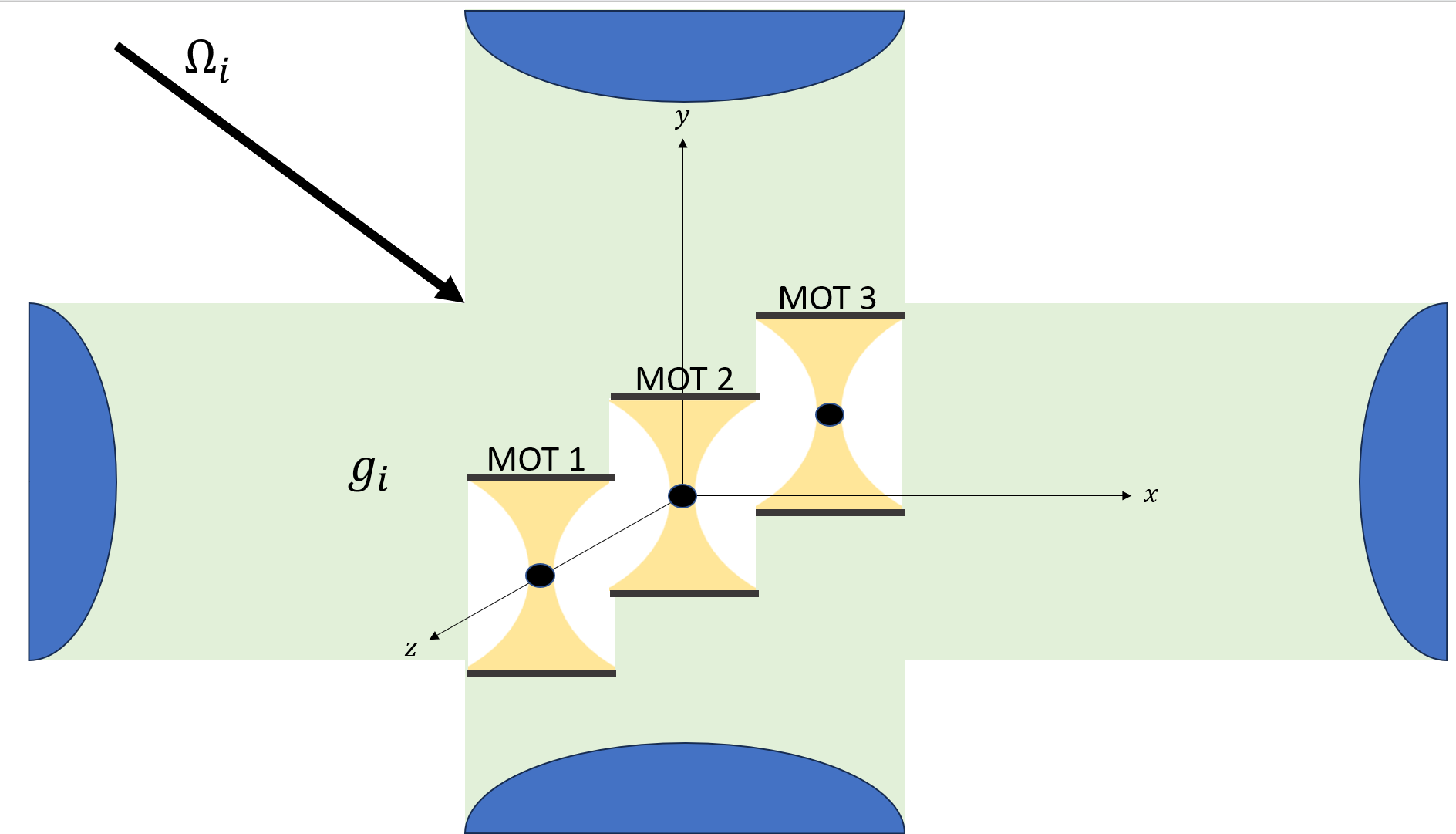}}
	
	\subfigure[]{\includegraphics[width=0.9\columnwidth]{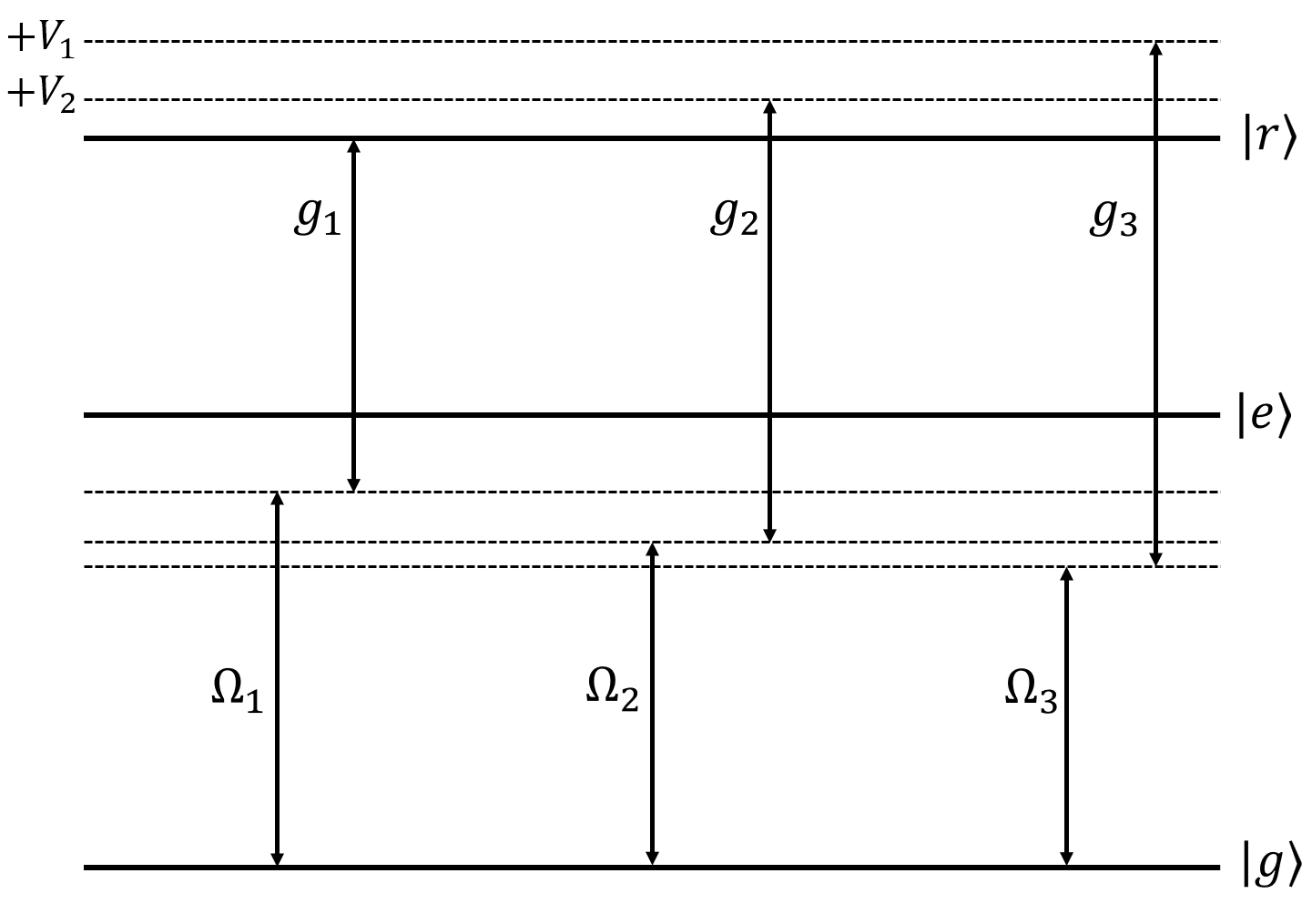}}
	\caption{Alternate implementation scheme for entanglement transfer using two-photon transitions with coupling to a multi-mode cavity, with coupling rates $g_i(t)$ and Gaussian pulses with Rabi frequencies $\Omega_i(t)$. a) Simplied scheme of setup with multi-mode cavity (mirrors-blue and spatial modes- green) and three atoms in magneto-optic traps. b) Diagram of three-level atom with ground and Rydberg levels and the intermediate state $\ket{e}$. One-photon detunings are all non-zero and large to ensure only two-photon transitions occur.}\label{fig: AltImp}
\end{figure}

We also propose an alternative strategy which takes advantage of two-photon transitions involving a virtual off-resonantly excited state. In this strategy, the cavity modes aren't chirped or spatially modulated. Instead the chirping of cavity modes and modulation of the coupling rates is created through the use of chirped pulses with Rabi frequencies $\Omega_{i}(t)=\Omega_{i,0}e^{-(t-t_i)^2/(2\tau_i^2)}$, and optical frequencies $\omega_{L,i}(t)=\omega_{ge}-\Delta_i+\alpha_{i,\Omega}(t)$. The cavity modes $j$ with coupling rates $g_j$ have frequencies $\omega_{er}+\Delta_j$. The three atoms are placed in magneto-optical traps located inside a multi-mode cavity with multiple optical axes, within the spot size of a pulsed laser, see Fig. \ref{fig: AltImp}. Both the pulses and the cavity modes are highly off-resonant with the two one-photon transitions, $\ket{g}\rightarrow\ket{e}$ and $\ket{e}\rightarrow\ket{r}$, but altogether satisfy the two-photon transition resonance condition , transferring population to $\ket{r}$ while preventing population of $\ket{e}$ for atomic transitions shifted by Rydberg-Rydberg interactions, as seen in Fig. \ref{fig: AltImp} b). The effective two-photon transition coupling rates and detunings are then given by $g_{i,\text{eff}}(t)=\dfrac{g_i\Omega_i(t)}{\Delta_i}$ and $\Delta_i^{\text{eff}}(t)=\alpha_{i,\Omega}(t)+\dfrac{\abs{g_i}^2-\abs{\Omega_i(t)}^2}{2\Delta_i}$ respectively. 

Unlike the previous strategy where the chirp rates are dependent on the cavity geometry and mode numbers, this method gives significantly more freedom in choosing the chirp rates as well as the possibility of using non-linear chirp functions that would be more difficult to implement in the moving mirror cavity. This yields more flexibility in dealing with the complex light shifts that appear in high-dimensional STIRAP. The dependence of the effective coupling on the pulse shape also gives more freedom with the coupling rate modulation. This strategy is compatible with both local manipulation of individual atoms \cite{Bluvstein2022} in the array as well as global universal manipulation \cite{Cesa2023}. A limitation of this strategy would be a possibly reduced speed of the entanglement transfer process due to the dependence on the four photon transitions involving 2 cavity mode photons and 2 pulse photons for the 0P to 2P state transition.

The methods described here would be of great interest for experimentalists looking to apply the protocols developed in this work. While we have not considered noise and dissipation effects in this work, processes such as cavity leaking, atomic decay and stochastic fluctuations in the coupling rates and mode frequencies will generally lower the fidelity and adjustments must be made accordingly to counterbalance the effects of noise. A way to reduce the detrimental effects of noise is to continuously monitor the cavity with an auxiliary system that's dispersively coupled to the cavity and apply unitary operations to correct disturbances to the cavity trajectory in parameter space \cite{harraz2019optimal}. The investigation of the effectiveness of this method will be the subject of a sequel work.

\section{Conclusion}\label{Sec5Conc}
In this work, we have successfully developed two protocols to transfer entanglement from a Rydberg three-atom W state to a three-mode photonic state. In Sec. \ref{Sec5Prob}, we outlined the basic problem statement 
as well as gave our approach to achieve entanglement transfer through chirping of the cavity modes and mapping of these modes to  
transitions between three-atom states. We introduced the two-stage scheme that we use for our two protocols for entanglement transfer. 
In Sec. \ref{Sec5Model}, we developed the basic theoretical model for the three Rydberg atoms interacting with a chirped multimode cavity and described the two-stage scheme in detail. 

In the next two sections, we described our two protocols to achieve entanglement transfer, a completely adiabatic protocol using STIRAP and FSTIRAP in Sec. \ref{Sec5CFSTIRAPad}, and a non-adiabatic protocol that used chirps to control the non-adiabatic transition probabilities in Sec. \ref{Sec5CFSTIRAPnonad}. 
Numerical simulations showed successful transfer of entanglement for generation of the three-photonic mode W-state for both protocols. Lastly, in Sec. \ref{Sec5Impl}, we introduced two strategies for realization of our protocols in an experimental setting. We showed that a 3D multimode cavity with a single moving mirror, along with atoms guided by optical tweezers, 
can be used to obtain the required Gaussian coupling rates and linear chirps. We also showed another strategy where we could make use of chirped pulses of light, in addition to a multi-mode cavity, to generate two-photon transitions between the three-atom states 
with coupling rates and detunings that inherit properties from both the cavity modes and the pulses.

The results of this work have significant utility in the development of quantum hybrid technologies that make use of programmable arrays of neutral atoms and photons for quantum computing and networking. The work also highlights the use of both adiabatic and non-adiabatic transitions to realize robust quantum operations within multipartite systems. Future goals of research would be to extend the protocol we have developed here for larger Rydberg atom superpositions, to calculate the non-adiabatic Stokes phase for various time-dependent couplings and chirps, and consider additional quantum control techniques that can be used to achieve high-fidelity transfer in the presence of noise. 

\begin{acknowledgments}
We acknowledge the support of the Office of Naval Research.
\end{acknowledgments}

\appendix

\section{Adiabatic elimination of intermediate states} \label{Sec5Adiab}
We redefine the mode frequencies such that $\omega_i(t)=\omega^{\prime}_i(t)+\Delta_i(t)$,\\ where $\abs{\omega^{\prime}_i(t)},\abs{g_i(t)}\ll \abs{\Delta_i(t)}$ for $t\leq T$. Here, $\Delta_i(t)$ are the one-photon detunings that allow us to adiabatically eliminate the 1P states. 

Starting from the field interaction Hamiltonian $H(t)$ \eqref{Hamil}, we attempt to find a computational basis that diagonalizes the inter-atomic interaction $V=\sum_{j_2>j_1}V_R(\vec{r}_{j_1},\vec{r}_{j_1})$. We take the operator basis $D^{n_1,n_2}$ that obeys,

\begin{eqnarray}
    V=(n_1V_1+n_2V_2)D^{n_1,n_2}
\end{eqnarray}

We want to expand $H(t)$ in this basis. We use the relation,

\begin{align}
\begin{split}
    &e^{iVt}\sigma_j^{\pm}e^{-iVt}=e^{iVt}\sigma_j^{\pm}\sum_{n_1,n_2}D^{n_1,n_2}e^{-iVt}\\
    &=\sum_{n_1,n_2}\left[\left(\sigma_{j+1}^{11}\sigma_{j-1}^{11}+e^{\pm iV_1t}(\sigma_{j+1}^{22}\sigma_{j-1}^{11}+\sigma_{j+1}^{11}\sigma_{j-1}^{22})\right.\right.\\
    &\left.\left.+e^{\pm i2V_1t}\sigma_{j+1}^{22}\sigma_{j-1}^{22}\right)\right.\\
    &\left.\left.\times(\sigma_{j+2}^{11}\sigma_{j-2}^{11}+e^{\pm iV_2t}(\sigma_{j+2}^{22}\sigma_{j-2}^{11}+\sigma_{j+2}^{11}\sigma_{j-2}^{22})\right.\right.\\
    &\left.\left.+e^{\pm i2V_2t}\sigma_{j+2}^{22}\sigma_{j-2}^{22}\right)\right]\sigma_j^{\pm}\sum_{n_1,n_2}D^{n_1,n_2}
\end{split}
\end{align}

Then all time-dependent terms in the Hamiltonian are only $c$-numbers. Returning to our adiabatic elimination procedure, we define $H(t)=H_c(t)+H_I(t)$, where $H_c(t)$ is the chirp term and $H_I(t)$ is the atom-cavity interaction, and use the below form of the Schrodinger equation,

\begin{align}
\begin{split}
    &\dot{\psi}(t)=-\dfrac{i}{\hbar}H_c(t)\psi(t)-\dfrac{i}{\hbar}H_I(t)\psi(t_0)\\
    &-\dfrac{1}{\hbar^2}\int_{t_0}^{t}dt^{\prime}\text{ }H_I(t)H_I(t^{\prime})\psi(t^{\prime})  
\end{split}
\end{align}

The second term purely contains fast terms of the form $e^{iwt}$ and can be ignored. The third term contains the two-photon resonant terms that non-trivially contribute. We have terms with form $a_j^{\dag}a_j$, $a_ja_j^{\dag}$ that represent the dynamic Stark shifts to the atom-cavity states and terms with form $a_j^{\dag}a_k^{\dag}$, $a_ka_j$ that represent the two-photon transitions between the $V_{0P}$ and $V_{2P}$ manifolds. Since we chose large 1P detunings, we use the weak-coupling approximation to substitute $\psi(t^{\prime})$ by $\psi(t)$ to bring the equation back to a time-local form. We determine a new time local Hamiltonian that only contains two-photon dynamics. For our 3 atom, 3 mode basis, we obtain the Stark shift Hamiltonian and two-photon absorption/emission Hamiltonians given by \eqref{HamilStark} and \eqref{HamilTwoPhoton} respectively. This treatment justifies the elimination of the $V_{1P}$ manifold.

For the remainder of this section, we describe the Hamiltonian for each subsystem $\mathcal{H}_{i,j}$ 
and derive the corresponding supereffective Hamiltonian $H_{i,j}^{\text{eff}}$ in the state basis, and forego use of the cavity operators $a_{k}$. 

\subsection{Stage 1}
The detunings of each cavity mode from the transition frequency of respective  atomic state are depicted in Fig. \ref{Diag}, where stage 1 concerns time evolution within interval $[t_0, t_1]$. The total wavefunction $\psi(t)$ for stage 1 is 
presented as the direct sum of three wave-functions $\psi_{1,1}(t)\oplus\psi_{1,2}(t)\oplus\psi_{1,3}(t)$, where $\psi_{1,i}(t)$ for $i=1,2$ correspond to the first and the second transfer process, defined below, and $\psi_{1,3}(t)$ is the wavefunction containing states that are unpopulated throughout stage 1, due to large one-photon detunings, forming the adiabatically eliminated subspace. Note that we do not include states which contribute to the Stark shifts but do not contribute to the transitions between 0P and 2P states. We instead insert the Stark shift components directly, using the Hamiltonian \eqref{HamilStark}, in the effective detunings.

The first and the second transfer  process require a complete population transfer from the 0P state $\ket{rgr,000}$ to the 2P state $\ket{ggg,110}$ and  a half population transfer from $\ket{rrg,000}$ and $\ket{ggr,000}$ to $\ket{ggg,101}$ respectively. We denote the first subsystem as $\mathcal{H}_{1,1}=\{\{\psi_{1,1;i}\}, H_{1,1}(t)\}$. For the first transfer process, the field interaction wavefunction $\psi_{1,1}(t)$ is given by, 
\begin{align}
	\psi_{1,1}(t)=
	e^{i\int d\tau\text{ }\omega_2^{\prime}(\tau)}\begin{pmatrix}
		c_{rgr,000}(t)\\
		c_{rgg,010}(t)\\
		c_{ggr,010}(t)\\
		c_{grg,010}(t)\\
		c_{ggg,110}(t)
	\end{pmatrix},\label{sys11basis}
\end{align}
and the Hamiltonian reads,
\begin{align}
	\begin{split}
		&H_{1,1}(t)=\\
		&\begin{pmatrix}
			-\omega_2^{\prime}(t) & g_{2}(t)  & g_{2}(t) & 0 & 0\\
			g_{2}(t) & \Delta_2(t)  & 0 & 0 & g_{1}(t)\\
			g_{2}(t) & 0  & \Delta_2(t) & 0 & g_{1}(t)\\
			0 & 0  & 0 & \Delta_2(t) & g_{1}(t)\\
			0 & g_{1}(t)  & g_{1}(t) & g_{1}(t) & \omega_1^{\prime}(t)+\delta_1(t)
		\end{pmatrix},    
	\end{split}
\end{align}
where $c_{rgr,000}(t_0)=1/\sqrt{3}$, all other probability amplitudes at $t_0$ are zero, and $\delta_1(t)=\Delta_1(t)+\Delta_2(t)$. We choose $\Delta_2(t)=-\Delta_1(t)=\Delta$, satisfying the two-photon resonance condition $\delta_1(t)=0$. For a large $\Delta$, the time derivatives of the 1P states can be set to zero giving us the two-level super-effective Hamiltonian in the basis $\{\ket{rgr,000},\ket{ggg,110}\}$- which reads,
\begin{equation}
	\begin{split}
		&H^{\text{eff}}_{1,1}(t)=\begin{pmatrix}
			-\Delta^{\text{eff}}_{1,12}(t)+\Delta^{0}_{1,12}(t) & g^{\text{eff}}_{1}(t)\\
			g^{\text{eff}}_{1}(t) &  \Delta^{\text{eff}}_{1,12}(t)+\Delta^{0}_{1,12}(t) 
		\end{pmatrix},    
	\end{split}\label{Ham2}
\end{equation}
where the terms are described below,
\begin{equation}
	\begin{split}
		&g^{\text{eff}}_{1,12}(t)=\dfrac{2g_{1}(t)g_{2}(t)}{\Delta},\\
		&\Delta^{\text{eff}}_{1,12}(t)=\dfrac{\omega_1^{\prime}(t)+\omega_2^{\prime}(t)}{2}-\dfrac{1}{2}\left(\braket{rgr,000}{H_S|rgr,000}\right.\\
        &\left.-\braket{ggg,110}{H_S|ggg,110}\right),\\
		&\Delta^{0}_{1,12}(t)=\dfrac{-\omega_1^{\prime}(t)+\omega_2^{\prime}(t)}{2}+\dfrac{1}{2}\left(\braket{rgr,000}{H_S|rgr,000}\right.\\
        &\left.+\braket{ggg,110}{H_S|ggg,110}\right).
	\end{split}\label{Ham2par}
\end{equation}
We give a similar treatment for the second transfer process. We denote the second subsystem of states and Hamiltonian as $\mathcal{H}_{1,2}=\{\{\psi_{1,2;i}\}, H_{1,2}(t)\}$.The field interaction wavefunction $\psi_{1,2}(t)$ is given by,
\begin{equation}
	\psi_{1,2}(t)(t)=
	e^{i\int d\tau\text{ }\omega_3^{\prime}(\tau)}\begin{pmatrix}
		c_{grr,000}(t)\\
		c_{rrg,000}(t)\\
		c_{ggr,001}(t)\\
		c_{grg,001}(t)\\
		c_{rgg,001}(t)\\
		c_{ggg,101}(t)
	\end{pmatrix},\label{sys12basis}
\end{equation}
where the Hamiltonian reads,
\begin{equation}
	\begin{split}
		&H_{1,2}(t)=\\
		&\begin{pmatrix}
			-\omega_3^{\prime}(t) & 0  & g_{3}(t) & g_{3}(t) & 0 & 0\\
			0 & -\omega_3^{\prime}(t)  & 0 & g_{3}(t) & g_{3}(t) & 0\\
			g_{3}(t) & 0  & \Delta_3(t) & 0 & 0 & g_{1}(t)\\
			g_{3}(t) & g_{3}(t)  & 0 & \Delta_3(t) & 0 & g_{1}(t)\\
			0 & g_{3}(t)  & 0 & 0 & \Delta_3(t) & g_{1}(t)\\
			0 & 0 & g_{1}(t)  & g_{1}(t) & g_{1}(t) & \tilde{\omega}(t)
		\end{pmatrix},    
	\end{split}\label{Ham31}
\end{equation}
where $c_{grr,000}(t_0)=c_{rrg,000}(t_0)=1/\sqrt{3}$, all other probability amplitudes are zero at $t_0$, $\tilde{\omega}(t)=\omega_1^{\prime}(t)+\delta_2(t)$ and $\delta_2(t)=\Delta_1(t)+\Delta_3(t)$. We set $\Delta_3(t)=-\Delta_1(t)=\Delta$, satisfying the two-photon resonance condition $\delta_2(t)=0$. Adiabatic elimination of the 1P states gives us the three-level super-effective Hamiltonian in the basis $\{\ket{rrg,000},\ket{grr,000},\ket{ggg,101}\}$- which reads,

\begin{equation}
	\begin{split}
		&H^{\text{eff}}_{1,2}(t)=\\
		&\begin{pmatrix}
			\tilde{\Delta}_{1,1}(t) & g^{\text{eff}}_{1,33}(t) & g^{\text{eff}}_{1,13}(t)\\
			g^{\text{eff}}_{1,33}(t) & \tilde{\Delta}_{1,1}(t) & g^{\text{eff}}_{1,13}(t)\\
			g^{\text{eff}}_{1,13}(t) & g^{\text{eff}}_{1,13}(t) &  \tilde{\Delta}_{1,2}(t) 
		\end{pmatrix},    
	\end{split}
\end{equation}
where the terms are given below,
\begin{equation}
	\begin{split}
		&\tilde{\Delta}_{1,1}(t)= -\Delta^{\text{eff}}_{1,13}(t)+\Delta^{0}_{1,13}(t),\\
		&\tilde{\Delta}_{1,2}(t)= \Delta^{\text{eff}}_{1,13}(t)+\Delta^{0}_{1,13}(t),\\     
		&g^{\text{eff}}_{1,33}(t)=\dfrac{g_{3}(t)g_{3}(t)}{\Delta},\\
		&g^{\text{eff}}_{1,13}(t)=\dfrac{2g_{1}(t)g_{3}(t)}{\Delta},\\
		&\Delta^{\text{eff}}_{1,13}(t)=\dfrac{\omega_1^{\prime}(t)+\omega_3^{\prime}(t)}{2}-\dfrac{1}{2}\left(\braket{rrg,000}{H_S|rrg,000}\right.\\
        &\left.-\braket{ggg,101}{H_S|ggg,101}\right),\\
		&\Delta^{0}_{1,13}(t)=\dfrac{-\omega_1^{\prime}(t)+\omega_3^{\prime}(t)}{2}+\dfrac{1}{2}\left(\braket{rrg,000}{H_S|rrg,000}\right.\\
        &\left.+\braket{ggg,101}{H_S|ggg,101}\right).
	\end{split}\label{THSterms}
\end{equation}

\begin{figure}[h!]
	\centering	\includegraphics[width=0.6\columnwidth]{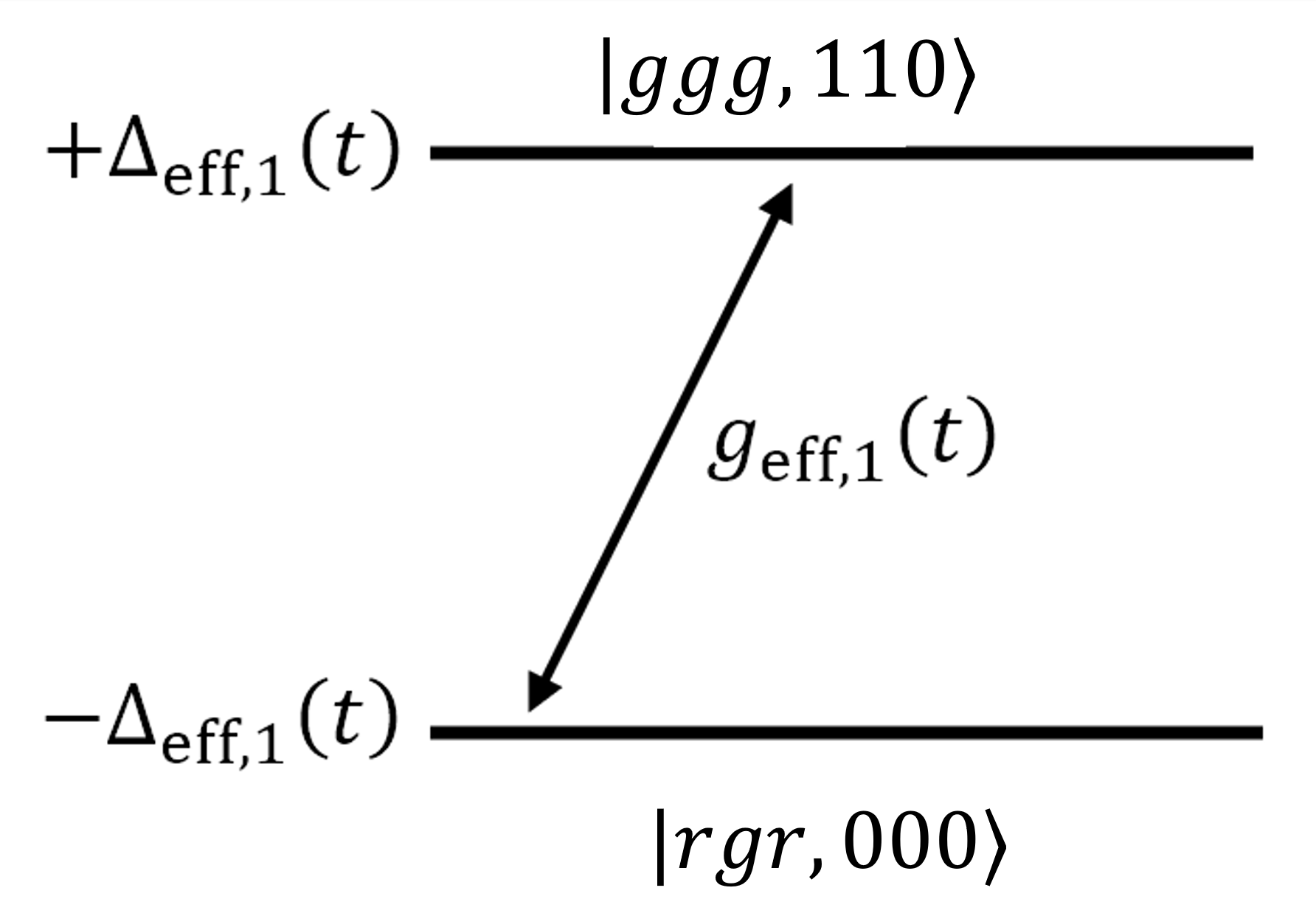}
	\caption{The two level super-effective system.}\label{TLSS}
\end{figure}

\begin{figure}[h!]
	\centering
	\includegraphics[width=0.9\columnwidth]{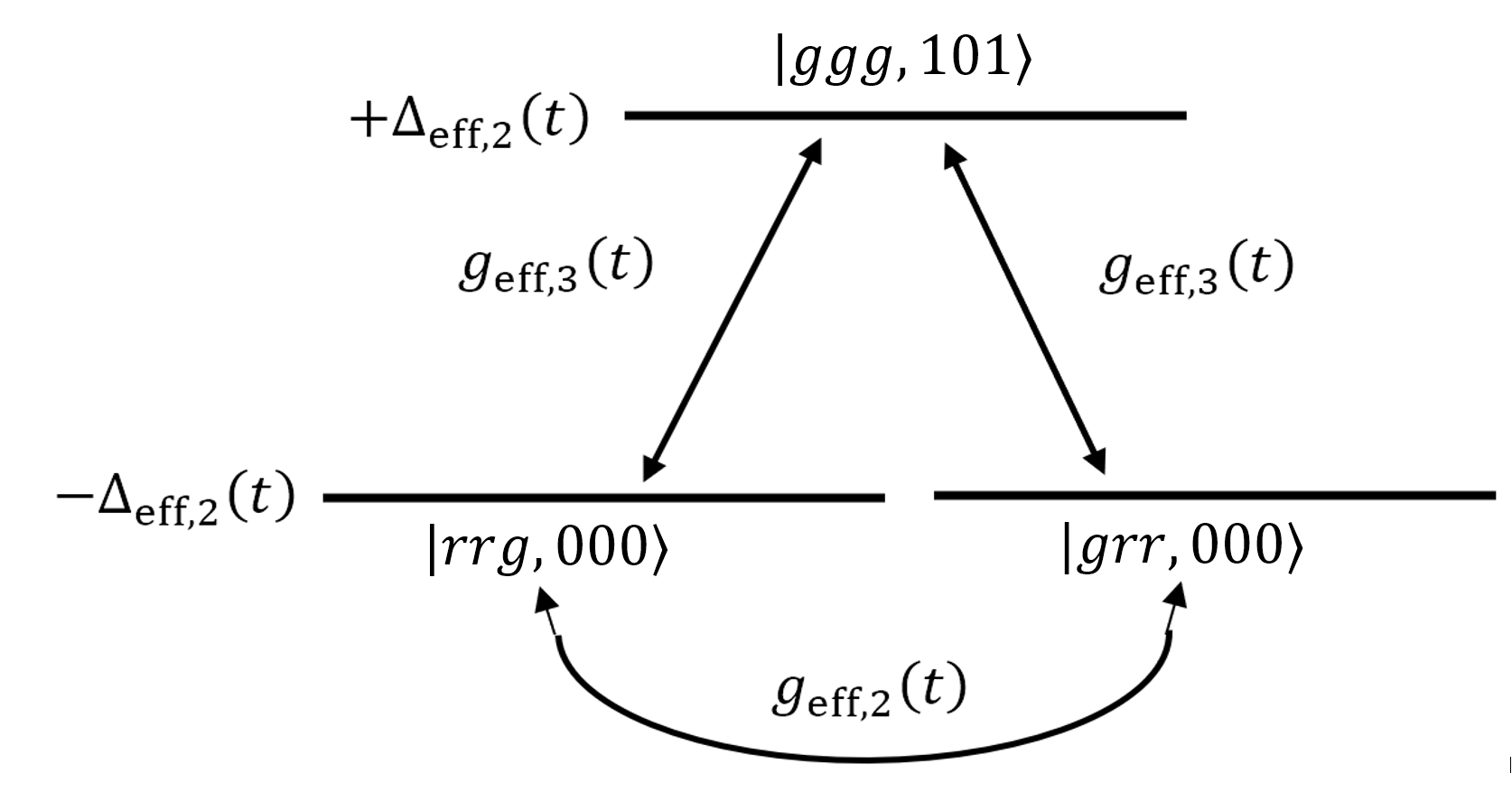}
	\caption{The three level super-effective system.}\label{ThLSS}
\end{figure}

Figures \ref{TLSS} and \ref{ThLSS} show the super-effective systems for each transfer process within stage 1. The above concludes our derivation of the adiabatic elimination part for stage 1. The dynamics for stage 2 processes is considered next. Between the two stages, the total  wavefunction evolves adiabatically for a duration $t=t_2-t_1$. This can be used to set the local phases in the wavefunction before stage 2 begins. 

\subsection{Stage 2}
The detunings of each cavity mode with each transition between the atom-cavity states follow Fig. \ref{Diag} where stage 2 concerns time evolution during the interval $[t_2,t_3]$.

In a similar treatment as with stage 1, $\psi(t)$ for stage 2 is decoupled as the direct sum $\psi_{2,1}(t)\oplus\psi_{2,2}(t)\oplus\psi_{2,3}(t)$ where $\psi_{2,i}(t)$ for $i=1,2$ corresponds to the first and the second transfer process described below and $\psi_{2,3}(t)$ is the remainder wavefunction similarly defined as in stage 1.

For the first transfer process, the objective is no population transfer from the $\ket{ggg,110}$ and $\ket{ggg,101}$ states, and $\psi_{2,1}(t)$ should be time-independent apart from a global phase. Since modes 1 and 2 are detuned sufficiently far away from the $\ket{rgr,000}\rightarrow\ket{rgg,100},\ket{ggr,100}$ and $\ket{rgr,000}\rightarrow\ket{rgg,010},\ket{ggr,010}$ transitions, and the two-photon detunings are non-zero, there is no pathway back to the 0P state manifold.  The first subsystem of states is given by $\mathcal{H}_{2,1}=\{\{\psi_{2,1;i}\}, H_{2,1}(t)\}$. The 
wavefunction for the first process reads as,

\begin{align}
	\psi_{2,1}(t)=
	e^{i\int_{0}^{t} d t^{\prime}\text{ }\omega_1^{\prime}(t^{\prime})}\begin{pmatrix}
		c_{ggg,101}(t)\\
		c_{ggg,110}(t)\\
		c_{rgg,100}(t)\\
		c_{ggr,100}(t)\\
		c_{grg,100}(t)
	\end{pmatrix},
\end{align}
with Hamiltonian given below,
\begin{align}
	\begin{split}
		&H_{2,1}(t)=\\
		&\begin{pmatrix}
			\omega_3^{\prime}(t) & 0 & g_{3}(t)  & g_{3}(t) & g_{3}(t)\\
			0 & \omega_2^{\prime}(t) & g_{2}(t)  & g_{2}(t) & g_{2}(t)\\
			g_{3}(t) & g_{2}(t) & \Delta_2(t)  & 0 & 0\\
			g_{3}(t) & g_{2}(t) & 0  & \Delta_2(t) & 0\\
			g_{3}(t) & g_{2}(t) & 0  & 0 & \Delta_2(t)
		\end{pmatrix},    
	\end{split}
\end{align}
where $c_{ggg,110}(t_2)=1/\sqrt{3}$, $c_{ggg,101}(t_2)=1/\sqrt{3}$, and all other probability amplitudes are zero at $t_2$. Setting $\abs{\Delta_2(t)}=\abs{\Delta_2}\gg \abs{g_2(t)}, \abs{g_3(t)}$ allows us to adiabatically eliminate the 1P states giving us adiabatic evolution of the amplitudes $c_{ggg,110}(t)$ and $c_{ggg,101}(t)$,

\begin{equation}
	\begin{split}
		\dot{c}_{ggg,101}(t)=-i\left(\omega_3^{\prime}(t)+\braket{ggg,101}{H_S|ggg,101}\right)c_{ggg,101}(t)\\
		\dot{c}_{ggg,110}(t)=-i\left(\omega_2^{\prime}(t)+\braket{ggg,110}{H_S|ggg,110}\right)c_{ggg,110}(t).\label{1Devo}    
	\end{split}
\end{equation}

For the second transfer process, the evolution is similar to the first stage counterpart involving the supereffective three-level system, except for the requirement that we need a complete transfer of remaining 1/3 of population from states $\ket{rrg,000}$ and $\ket{grr,000}$ to $\ket{ggg,011}$. The second subsystem of states is given by $\mathcal{H}_{2,2}=\{\{\psi_{2,2;i}\}, H_{2,2}(t)\}$. The field interaction wavefunction $\psi(t)$ is, 
\begin{equation}
	\psi_{2,2}(t)=
	e^{i\int_0^t dt^{\prime}\text{ }\omega_3^{\prime}(t^{\prime})}\begin{pmatrix}
		c_{grr,000}(t)\\
		c_{rrg,000}(t)\\
		c_{ggr,001}(t)\\
		c_{grg,001}(t)\\
		c_{rgg,001}(t)\\
		c_{ggg,011}(t)
	\end{pmatrix}\label{sys22basis}
\end{equation}
and the Hamiltonian reads,
\begin{equation}
	\begin{split}
		&H_{2,2}(t)=\\
		&\begin{pmatrix}
			-\omega_3^{\prime}(t) & 0  & g_{3}(t) & g_{3}(t) & 0 & 0\\
			0 & -\omega_3^{\prime}(t)  & 0 & g_{3}(t) & g_{3}(t) & 0\\
			g_{3}(t) & 0  & \Delta_3(t) & 0 & 0 & g_{2}(t)\\
			g_{3}(t) & g_{3}(t)  & 0 & \Delta_3(t) & 0 & g_{2}(t)\\
			0 & g_{3}(t)  & 0 & 0 & \Delta_3(t) & g_{2}(t)\\
			0 & 0 & g_{2}(t)  & g_{2}(t) & g_{2}(t) & \tilde{\omega}_2(t)
		\end{pmatrix},    
	\end{split}
\end{equation}
where $c_{grr,000}(t_2)=c_{rrg,000}(t_2)=1/\sqrt{6}$, all other probablity amplitudes are zero at $t_2$,   $\tilde{\omega}_2(t)=\omega_2^{\prime}(t)+\delta_3(t)$, and $\delta_3(t)=\Delta_2(t)+\Delta_3(t)$. We set $\Delta_3(t)=-\Delta_2(t)=\Delta_2$, satisfying the two-photon resonance condition $\delta_3(t)=0$. Adiabatic elimination of the 1P states gives us the  three-level super-effective Hamiltonian in the basis $\{\ket{rrg,000},\ket{grr,000},\ket{ggg,011}\}$, which reads,

\begin{equation}
	\begin{split}
		&H^{\text{eff}}_{2,2}(t)=\\
		&\begin{pmatrix}
			\tilde{\Delta}_{2,1}(t) & g^{\text{eff}}_{2,33}(t) & g^{\text{eff}}_{2,23}(t)\\
			g^{\text{eff}}_{2,33}(t) & \tilde{\Delta}_{2,1}(t) & g^{\text{eff}}_{2,23}(t)\\
			g^{\text{eff}}_{2,23}(t) & g^{\text{eff}}_{2,23}(t) &  \tilde{\Delta}_{2,2}(t) 
		\end{pmatrix},    
	\end{split}
\end{equation}
with terms given below,
\begin{equation}
	\begin{split}
		&\tilde{\Delta}_{2,1}(t)= -\Delta^{\text{eff}}_{2,23}(t)+\Delta^{0}_{2,23}(t),\\
		&\tilde{\Delta}_{2,2}(t)= \Delta^{\text{eff}}_{2,23}(t)+\Delta^{0}_{2,23}(t),\\     
		&g^{\text{eff}}_{2,33}(t)=\dfrac{g_{3}(t)g_{3}(t)}{\Delta},\\
		&g^{\text{eff}}_{2,23}(t)=\dfrac{2g_{2}(t)g_{3}(t)}{\Delta},\\
		&\Delta^{\text{eff}}_{2,23}(t)=\dfrac{\omega_1^{\prime}(t)+\omega_3^{\prime}(t)}{2}-\dfrac{1}{2}\left(\braket{rrg,000}{H_S|rrg,000}\right.\\
        &\left.-\braket{ggg,011}{H_S|ggg,011}\right),\\
		&\Delta^{0}_{2,23}(t)=\dfrac{-\omega_1^{\prime}(t)+\omega_3^{\prime}(t)}{2}+\dfrac{1}{2}\left(\braket{rrg,000}{H_S|rrg,000}\right.\\
        &\left.+\braket{ggg,011}{H_S|ggg,011}\right).
	\end{split}
\end{equation}

We have greatly simplified our original 17-state problem into a sequence of problems with lower-dimensional two- and three-level super-efficient systems. By careful choice of coupling rates $g_i(t)$ and frequencies $\omega_i(t)$ we can control each population transfer process with diabatic and adiabatic transitions within a single avoided crossing. The method we use to capitalize on this will be based on the adiabatic-impulse (or transfer-matrix) approximation. In the next section, we give a rigorous study of two- and three-level super-efficient systems using the formalism of diabatic and adiabatic states at avoided crossings, deriving the diabatic and adiabatic states for each $\mathcal{H}_{i,j}$, so as to build up to the final results that describe the time evolution corresponding to the subsystem. We furthermore use the Morris-Shore transformations to decouple dark states from the super-effective three level subsystems so that we can simplify even further to super-effective two-level systems. 

\begin{figure}[h!]
	\centering
	\includegraphics[width=0.9\columnwidth]{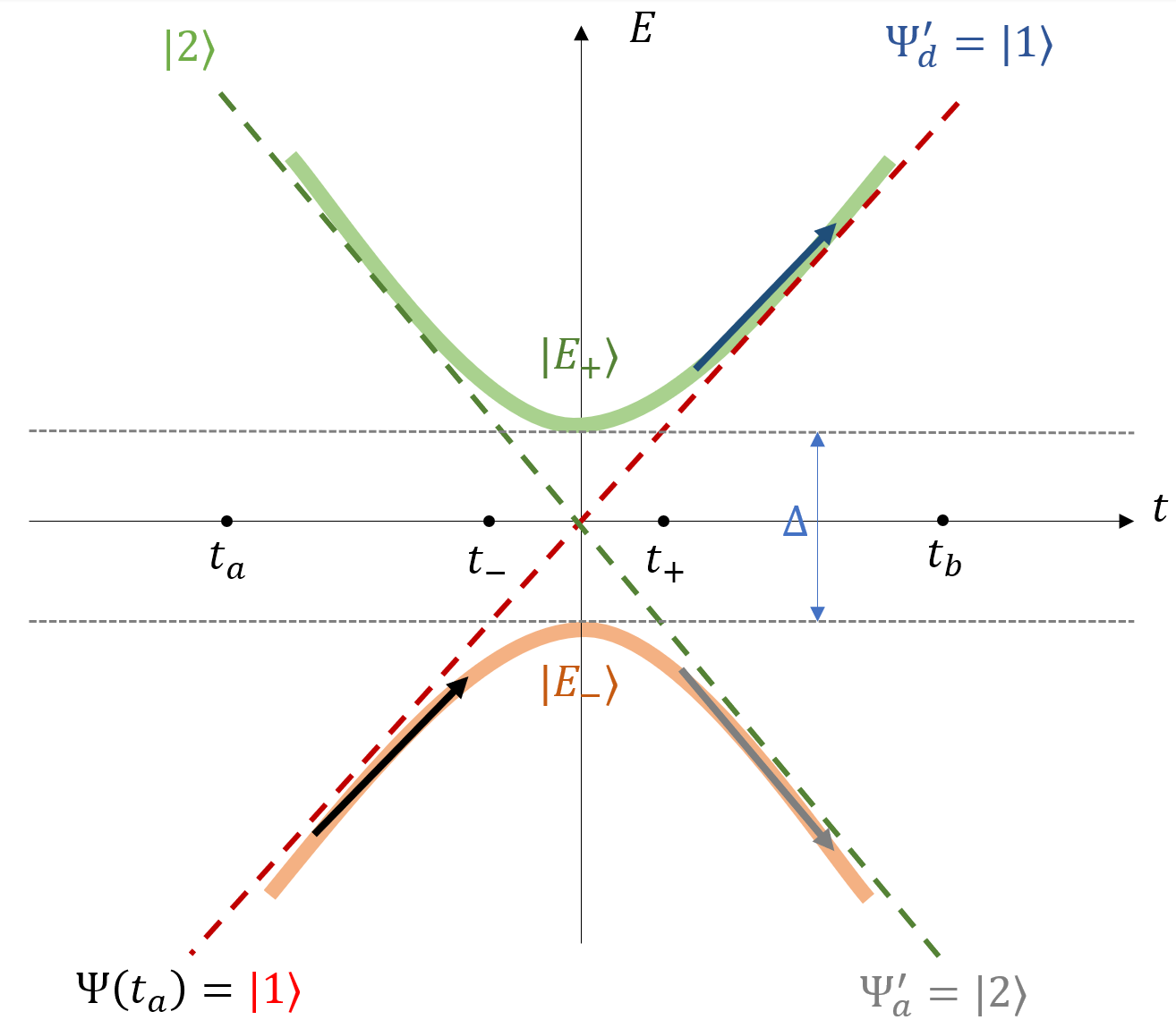}
	\caption{Evolution of the wave function near an avoided crossing between two states. Green/red dashed lines are diabatic states ($\ket{1}$, $\ket{2}$) and solid lines are adiabatic states ($\ket{E_{-}}$, $\ket{E_{+}}$). The time interval $[t_a, t_b]$ is segmented by the sub-interval $[t_{-}, t_{+}]$ where there is a sharp non-adiabatic transition. Red/green refers to states that start in the ground/excited state manifold. The adiabatic/diabatic trajectories, represented by gray and dark blue arrows, terminate at $\Psi_a^{\prime}$/$\Psi_d^{\prime}$ respectively. The probability of diabatic passage depends on the energy gap, $\Delta$, and the slope of the adiabatic states involved at the avoided crossing.} \label{fig: LZmodel}
\end{figure}

\section{The adiabatic and diabatic states}\label{Sec5Adiabdiab}
For each subsystem $\mathcal{H}_{i,j}$, we define the diabatic states as the bare states of the super-effective Hamiltonian $H_{i,j}^{\text{eff}}$, with the energies given by the diagonal entries of the Hamiltonian, while the adiabatic states diagonalize $H_{i,j}^{\text{eff}}$. 
As an example, Fig. \ref{fig: LZmodel} shows adiabatic (solid) and diabatic (dashed) states of a two-level system at an avoided crossing.

We first study the supereffective TLS of system $\mathcal{H}_{1,1}$, described by Eq. \eqref{Ham2}, and apply transformation $R_{0}(t)=e^{-i\int_0^t dt^{\prime} \Delta^0_{1,13}(t^{\prime})}$. Using the terms in \eqref{Ham2par}, we define the adiabatic energy $A_1(t)=\sqrt{\Delta^{\text{eff}}_{1,13}(t)^2+\abs{g^{\text{eff}}_{1,13}(t)}^2}$, and the mixing angle implicitly given by $\tan 2\varphi_1(t)=\dfrac{\abs{g^{\text{eff}}_{1,13}(t)}}{\Delta^{\text{eff}}_{1,13}(t)}$- which are part of the parametric vector $\lambda_1(t)=(A_1(t),\varphi_1(t))$ . The diabatic basis Hamiltonian in parameter space is then given by,
\begin{equation}
	\begin{split}
		&H^{\text{di}}_{1,1}(\lambda_1)=-A_1\left(\cos 2\varphi_1\hat{\sigma}_z-\sin 2\varphi_1\hat{\sigma}_x\right)
	\end{split}\label{Ham10}
\end{equation}
Where the diabatic state basis, in terms of the basis given in \eqref{sys11basis} for $\mathcal{H}_{1,2}$, and diabatic state energies are given by,
\begin{equation}
	\begin{split}
		&D_{1/2}(\lambda_1)=\mp A_1\cos 2\varphi_1\\
		&\ket{1}=e^{i\int_0^t dt^{\prime} \Delta^0_{1,13}(t^{\prime})}\ket{rgr,000}\\
		&\ket{2}=e^{i\int_0^t dt^{\prime} \Delta^0_{1,13}(t^{\prime})}\ket{ggg,110}
	\end{split}
\end{equation}

By diagonalizing the Hamiltonian, we obtain the adiabatic energies and states,
\begin{equation}
	\begin{split}
		&E_{\pm}(\lambda_1)=\pm A_1\\
		&\ket{w_{+}(\lambda_1)}=\sin\varphi_1\ket{1}+\cos \varphi_1\ket{2}\\
		&\ket{w_{-}(\lambda_1)}=\cos\varphi_1\ket{1}-\sin \varphi_1\ket{2}
	\end{split}\label{adiabs22}
\end{equation}
With the Hamiltonian in the adiabatic basis defined as below,
\begin{equation}
	\begin{split}
		&H_{1,1}^{\text{ad}}(\lambda_1)=-A_1\ketbra{w_-}{w_-}+A_1\ketbra{w_+}{w_+}+\\
  &i\dot{\varphi}_1\left(\ketbra{w_+}{w_-}-\ketbra{w_-}{w_+}\right)
	\end{split}\label{Ham11}
\end{equation}

For complete population transfer, we can ensure this occurs adiabatically by introducing a pulse shaped coupling for $g^{\text{eff}}_{1}(t)$ while sweeping $\Delta^{\text{eff}}_{1}$ from positive to negative, thereby tracing a path $\lambda_1(t)$ in parameter space where $2\varphi_1(\lambda_1): 0\rightarrow \pi$. On the path $\lambda_1(t)$, we satisfy the condition $\zeta\ll 1$ where $\zeta=\abs{\dfrac{\dot{\varphi}_1}{2A}}$ is the adiabatic parameter. Assuming that the system starts in state $\ket{w_-(t)}$, $\psi(t_0)=\ket{w_-(0)}$, the wavefunction for this system is given by,

\begin{equation}
	\psi(t)=e^{-i\int_{t_0}^{t}dt^{\prime}\text{ }E_{-}(t^{\prime})}e^{\int_{S(\lambda_1)} dS\text{ }\Braket{w_{-}|\nabla_S |w_{-}}}\ket{w_{-}(t)}\label{adiev}
\end{equation}

The above equation gives the wavefunction of the super-effective TLS for an adiabatic evolution. Assuming we choose a path $\lambda_1(t)$ where the adiabatic condition is satisfied, and the adiabatic state $\ket{w_-(t)}$ satisfies $\lim_{t\rightarrow t_0}\ket{w_-(t)}=\ket{1}$ and $\lim_{t\rightarrow t_1}\ket{w_-(t)}=\ket{2}$, we end up with complete population in $\ket{2}$ with an acquired dynamical phase $\int_{t_0}^{t_1}dt^{\prime}\text{ }E_{-}(t^{\prime})$. The STIRAP method provides us with such paths $\lambda_1$ that satisfy the mixing angle condition $\lim_{t\rightarrow+\infty}\varphi_1(\lambda_1)=\pi/2$ condition. This completes the required evolution for system $\mathcal{H}_{1,1}$. 
We note that we can use adiabatic evolution to also end up in a superposition of the two diabatic states if we instead choose the adiabatic path $\lambda_2(t)$ where $2\varphi_1(\lambda_2): 0\rightarrow \pi/2$. Such an evolution is satisfied by the FSTIRAP protocol and is possible when the chirp rates are equal and opposite- resulting in the diabatic states having equal energy in the infinite past and future.

The same analysis can be extended to the one-level super-effective system of $\mathcal{H}_{2,1}$, with dynamical phase given by integrating Eq. \eqref{1Devo} from $t=t_2$ to $t=t_3$.

For the the super-effective three level subsystems of $\mathcal{H}_{1,2}$ and $\mathcal{H}_{2,2}$, we require some additional work before we derive the adiabatic and diabatic state basis and Hamiltonians. Since these two subsystems are equivalent within a change of parameters, we focus on $\mathcal{H}_{1,2}$ and extend the later derived results to $\mathcal{H}_{2,2}$. 

We first apply transformation $P_{0}(t)=e^{-i\int_0^t dt^{\prime} \Delta^0_{1,13}(t^{\prime})}$. We find that the degeneracy of the two lower levels is broken due to the effective coupling $g^{\text{eff}}_{1,33}(t)$. We can then apply the Morris-Shore transformations \cite{Zlatanov2020} to simplify the three level system with two degenerate levels to a set of an uncoupled one- and two-level systems. 

Assuming that $g^{\text{eff}}_{1,33}(t)$ is non-negative, the transformation $P_1(t)$ is introduced where,

\begin{equation}
	\begin{split}
		&P_1(t)=\exp\left(-\dfrac{i}{2}\int_0^t dt^{\prime}\text{ }g^{\text{eff}}_{1,33}(t^{\prime})\begin{pmatrix}
			-1 & 2 & 0\\
			2 & -1 & 0\\
			0 & 0 & 1
		\end{pmatrix}\right),
	\end{split}
\end{equation}
Where $\tilde{\Delta}^{\text{eff}}_{1,13}(t)={\Delta}^{\text{eff}}_{1,13}(t)+g^{\text{eff}}_{1,33}(t)/2$. We introduce a further transformation $P_2(t)$ that decouples a state from the system,
\begin{equation}
	\begin{split}
		&P_2=\dfrac{1}{\sqrt{2}}\begin{pmatrix}
			1 & 1 & 0\\
			-1 & 1 & 0\\
			0 & 0 & \sqrt{2}
		\end{pmatrix},
	\end{split}
\end{equation}
Giving us the Morris-Shore transformed effective Hamiltonian $H_{1,2}^{\text{eff,MS}}$,
\begin{equation}
	\begin{split}
		&H_{1,2}^{\text{eff,MS}}(t)=\\
		&\begin{pmatrix}
			-\tilde{\Delta}^{\text{eff}}_{1,13}(t) & 0 & 0\\
			0 & -\tilde{\Delta}^{\text{eff}}_{1,13}(t) & \sqrt{2}g^{\text{eff}}_{1,13}(t)\\
			0 & \sqrt{2}g^{\text{eff}}_{1,13}(t) &  \tilde{\Delta}^{\text{eff}}_{1,13}(t)
		\end{pmatrix},    
	\end{split}\label{THHam}
\end{equation}

The application of transformation $P_2P_1(t)P_0(t)$ yields the transformed diabatic basis $\{\ket{\tilde{0}},\ket{\tilde{1}},\ket{\tilde{2}}\}$. We see that $\ket{\tilde{0}}$ is uncoupled from the other two states, implying that it's a dark state. We proceed to derive the diabatic and adiabatic basis Hamiltonians explicitly. We define the adiabatic energy $A_2(t)=\sqrt{(\tilde{\Delta}^{\text{eff}}_{1,13}(t))^2+\abs{\sqrt{2}g^{\text{eff}}_{1,13}(t)}^2}$, and mixing angle $\tan 2\varphi_2(t)=\dfrac{\sqrt{2}\abs{g^{\text{eff}}_{1,13}(t)}}{\tilde{\Delta}^{\text{eff}}_{1,13}(t)}$, where the parametric vector is given by $\lambda_2(t)=(A_2(t),\varphi_2(t))$. The diabatic basis Hamiltonian in parameter space is given by,

\begin{equation}
	\begin{split}
		&H_{1,2}^{\text{di}}(\lambda_2)=-A_2\dfrac{\cos 2\varphi_2}{3}\left(2\sqrt{3}\hat{\sigma}_{z,2}+I\right)+A_2\sin 2\varphi_2\hat{\sigma}_{x,23}
	\end{split}\label{Ham12}
\end{equation}

where the $\hat{\sigma}$ matrices are the generalized Pauli matrices for $\mathrm{SU}(3)$ in terms of the diabatic state basis in the rotated frame $\{\ket{\tilde{0}},\ket{\tilde{1}},\ket{\tilde{2}}\}$. 
These states are given below in terms of the basis in \eqref{sys12basis} for $\mathcal{H}_{1,2}$,
\begin{equation}
	\begin{split}
		&D_{0}(\lambda_2)=-A_2\cos 2\varphi_2\\
		&D_{1,2}(\lambda_2)=\mp A_2\cos 2\varphi_2\\
		&\ket{\tilde{0}}=e^{i\int_0^t dt^{\prime} \left(\Delta^0_{1,13}(t^{\prime})-\frac{3g_{1,33}^{\text{eff}}(t^{\prime})}{2}\right)}\dfrac{\left(\ket{rrg,000}-\ket{grr,000}\right)}{\sqrt{2}}\\
		&\ket{\tilde{1}}=e^{i\int_0^t dt^{\prime} \left(\Delta^0_{1,13}(t^{\prime})+\frac{g_{1,33}^{\text{eff}}(t^{\prime})}{2}\right)}\dfrac{\left(\ket{rrg,000}+\ket{grr,000}\right)}{\sqrt{2}}\\
		&\ket{\tilde{2}}=e^{i\int_0^t dt^{\prime} \left(\Delta^0_{1,13}(t^{\prime})+\frac{g_{1,33}^{\text{eff}}(t^{\prime})}{2}\right)}\ket{ggg,101}
	\end{split}
\end{equation}
We diagonalize the Hamiltonian \eqref{THHam} to obtain the adiabatic states and energies,
\begin{equation}
	\begin{split}
		&E_0(\lambda_2), E_{\pm}(\lambda_2)=-A_2\cos 2\varphi_2, \pm A_2\\
		&\ket{w_0}=\ket{\tilde{0}}\\
		&\ket{w_{-}(\lambda_2)}=\cos\varphi_2\ket{\tilde{1}}+\sin\varphi_2\ket{\tilde{2}}\\
		&\ket{w_{+}(\lambda_2)}=\sin\varphi_2\ket{\tilde{1}}-\cos\varphi_2\ket{\tilde{2}}
	\end{split}
\end{equation}
Where the adiabatic basis Hamiltonian is given below,
\begin{equation}
	\begin{split}
		&H_{1,2}^{\text{ad}}(\lambda)=-A_2\ketbra{w_-}{w_-}+A_2\ketbra{w_+}{w_+}\\
		&-A_2\cos 2\theta_2\ketbra{w_0}{w_0}+i\dot{\varphi}_2\left(-\ketbra{w_-}{w_+}+\ketbra{w_+}{w_-}\right)
	\end{split}\label{Ham12adiab}
\end{equation}

We observe that $\ket{w_0}$ is uncoupled from the other two eigenstates, implying it's a dark state, and is furthermore equal to the diabatic state $\ket{\tilde{0}}$. The adiabatic evolution of the wavefunction is also given by Eq. \eqref{adiev}, after substituting $\lambda_1=\lambda_2$. Similar to the adiabatic evolution for $\mathcal{H}_{1,1}$, we can follow a path $\lambda_2(t)$ in parameter space where $2\varphi_2(\lambda_1): 0\rightarrow \pi$ to completely transfer population between the diabatic states $\ket{\tilde{1}}$ and $\ket{\tilde{2}}$. Extrapolating the just derived results to $\mathcal{H}_{2,2}$, we see that this path completes the required population transfer for the subsystem. For $\mathcal{H}_{1,2}$, in the case of equal and opposite chirp rates, we can achieve the required half population transfer adiabatically by using the FSTIRAP technique.

A complete description of adiabatic dynamics has been given for each transfer process in stage 1 and stage 2. However adiabatic population transfer in our model cannot be used to give us the half depopulation in the case of equal non-zero chirps. In this case, since the adiabatic state with energy $A(t)$ will necessarily converge to the diabatic state  with energy $D(t)$ satisfying the condition $\lim_{t\rightarrow +\infty}E(t)=D(t)$, (as seen in Fig. \ref{fig: LZmodel}), there is no mixing of the initial and final states. This is due to our choice of turning off the coupling rates $g_i(t)$ after each transfer process- which necessarily fixes the mixing angle $\varphi_2$ to equal integer multiples of $\pi/2$. We made this choice to avoid populating the unwanted 2P states. 
We proceed to develop a theoretical treatment of the non-adiabatic transition probabilities that accounts for the time dependence of the couplings $g_i(t)$ and mode frequencies $\omega_i(t)$.

\subsection{Non-adiabatic dynamics}

This subsection will concern the non-adiabatic evolution of subsystem $\mathcal{H}_{1,2}$ (and equivalently $\mathcal{H}_{2,2}$), as described by the adiabatic basis two-level Hamiltonian \eqref{Ham12adiab}, as we do not require non-adiabatic evolution for the other subsystems $\mathcal{H}_{i,1}$. We will work exclusively in the adiabatic basis \eqref{adiabs22}.

The time evolution of the wavefunction near an avoided crossing can be described by the adiabatic impulse approximation (AIA) \cite{Damski2006, Shevchenko2010} where we segment the time interval of evolution near the avoided crossing $[t_a,t_b]$ into sub-intervals $I_{-}=[t_a,t_{-}]$ and $I_{+}=[t_{+},t_b]$, where the dynamics are purely adiabatic, and interval $I_{d}=[t_{-},t_{+}]$ where there is a sharp non-adiabatic transition, see Fig. \ref{fig: LZmodel}. The time evolution operator during the whole interval is expressed as a product of unitary operators $U_{+}NU_{-}$ where $U_{\mp}$ are the unitary operators for adiabatic evolution during intervals $I_{\mp}$ and $N$ is the non-adiabatic transfer matrix for evolution during $I_{d}$. The adiabatic evolution operator for a time interval $[t_a,t_b]$, using Eq. \eqref{adiev}, is given by,

\begin{equation}
	U(t_a,t_b)=\exp(-i\int_{t_a}^{t_b}dt^{\prime}\text{ }(E_{+}(t^{\prime})-E_{-}(t^{\prime}))\sigma_z)
\end{equation}

And the non-adiabatic transfer matrix $N$ is given by,

\begin{equation}
	\begin{split}
		N=\begin{pmatrix}Re^{-i\phi_S} & -T\\
			T & Re^{i\phi_S}
		\end{pmatrix}    
	\end{split}
\end{equation}

where $R$, $T$ are called the reflection and transmission coefficients and $\phi_S$ is the Stokes phase \cite{Kayanuma1997}. An exact expression for the Stokes phase is given for the Landau-Zener model with constant coupling \cite{Shevchenko2010}, but it's more difficult to derive an exact expression in the case of time-dependent couplings.
The transition probability between the two adiabatic states is given by $\abs{T}^2$. We use the formalism developed by Dykhne, Davis and Pechukas \cite{Dykhne1962, Davis2008} for calculating the transition probability between adiabatic states by extending the Hamiltonian to the complex time plane. The contributions to the transmission coefficient comes from the classical turning points of the adiabatic energy difference $2A_2(t)$. At an avoided crossing, these turning points are located in the complex plane for time $z$ and we can lift the real time integral over $[t_{-},t_{+}]$ to a complex contour that circles these points.
Assuming that there are only two contributing turning points, $z=\pm z_0$, that both yield square root branch cuts on $A_2(z)$, the probability of a non-adiabatic transition is given by,

\begin{equation}
	\text{Pr}=\abs{T}^2=e^{-2\text{Im}\int_{-\infty}^{z_0}2A_2(z) dz},\label{Dykhneres}
\end{equation}
Where $z=z_0$ is the turning point in the upper half plane. This result, called the Dykhne formula, is a powerful tool for our purpose of determining parameters of the pulses and chirps that lead to a final state population.

\section{Derivation of non-adiabatic transition probabilities}\label{App5A}
We first introduce a time scaling $t\rightarrow\dfrac{t^{\prime}}{\epsilon}$ where $\epsilon>0$ defines the time scale over which time-dependent quantities in the Hamiltonian $H(\epsilon t)$ varies. We extend the time variable $t^{\prime}$ to the complex plane with complex variable $z=t+i\tau$ and define the complex phase function,

\begin{equation}
    D(z)=\int_{\gamma: z_1\rightarrow z}A(z) dz
\end{equation}

Assuming the wavefunction in the adiabatic basis is given by,
\begin{equation}
\begin{split}
    &\psi(z)=a_1(z)e^{-i(D(z)-D_0)}\ket{v_1(z)}\\
    &+a_2(z)e^{i(D(z)-D_0)}\ket{v_2(z)}
\end{split}
\end{equation}
It follows that the equations of motion of amplitudes $(a_1(z),a_2(z))$, with Hamiltonian $H(t)$, is given by,

\begin{equation}
    \begin{split}
        &\dfrac{d}{dz}\begin{pmatrix}
            a_1(z)\\
            a_2(z)
        \end{pmatrix}\\
        &=\begin{pmatrix}
            0 &-\dot{\theta}(z)e^{i(D(z)-D_0)}\\
            \dot{\theta}(z)e^{-i(D(z)-D_0)} & 0
        \end{pmatrix}
        \begin{pmatrix}
            a_1(z)\\
            a_2(z)
        \end{pmatrix},
    \end{split}
\end{equation}

Near the turning point, $A(z)=A^{(1)}(z_0)(z-z_0)^{1/2}+\mathcal{O}(z^{3/2})$ and $\dot{\theta}(z)=\dfrac{1}{4i(z-z_0)}+\mathcal{O}(z^{0})$. We segment the evolution contour, with respect to real time $t$ into intervals corresponding to adiabatic evolution, $t\in[-\infty,t_{-}]\cup[t_{+},\infty]$ and non-adiabatic evolution, $t\in [t_{-},t_{+}]$. Using the connection formula \cite{Dykhne1962} for eigenenergies and eigenstates at different times $z_1, z_2$ ,

\begin{align}
    \begin{split}
        &A(z_2)=\\
        &\sqrt{\left(A(z_1)+\dfrac{w_{11}(z_1,z_2)-w_{22}(z_1,z_2)}{2}\right)^2+\abs{w_{12}(z_1,z_2)}^2}\\
        &\ket{v_{\mp}(z_2)}=\dfrac{\sqrt{1\pm k(z_1,z_2)}\ket{v_{-}(z_1)}}{\sqrt{2}}\\
        &\mp\dfrac{\sqrt{1\mp k(z_1,z_2)}\ket{v_{+}(z_1)}}{\sqrt{2}}\\
        &w_{ij}(z_1,z_2)=\Braket{v_i(z_1)|H(z_2)-H(z_1)|v_j(z_1)}\\
        &k(z_1,z_2)=\dfrac{1}{A(z_2)}\left(A(z_1)+\dfrac{w_{11}(z_1,z_2)-w_{22}(z_1,z_2)}{2}\right)
    \end{split}
\end{align}
We ensure the continuity of the wavefunction at the endpoints $t=t_{\pm}$ shared by the adiabatic and non-adiabatic intervals. To obtain results of the non-adiabatic dynamics, we use the below transformation,
\begin{equation}
    b(z)=\dfrac{a_1(z)}{\sqrt{A(z)}}e^{-i/2(D(z)-D_0)}+\dfrac{ia_2(z)}{\sqrt{A(z)}}e^{i/2(D(z)-D_0)},
\end{equation}
To obtain an Airy differential equation,
\begin{equation}
    \dfrac{d^2b(z)}{dz^2}+\left(\dfrac{3A^{(1)}(z_0)}{4}\right)^2(z-z_0)b(z)=0.
\end{equation}

For $z\rightarrow -\infty$, we have,

\begin{equation}
    b(z)\sim a_1(-\infty)e^{-i/2(D(z)-D_0)}
\end{equation}

For $z\rightarrow +\infty$, we have,

\begin{equation}
    b(z)\sim a_1(+\infty)e^{-i/2(D(z)-D_0)}+ia_2(+\infty)e^{i/2(D(z)-D_0)}
\end{equation}

Assuming that there are only two turning points located above and below a point on the real time axis where $A=0$ gives a square root branch cut, the probability of a non-adiabatic transition is given by:

\begin{equation}
    \text{Pr}=\abs{a_2(+\infty)}^2=e^{-2\text{Im}\int_{-\infty}^{z_0}2A(z) dz},
\end{equation}
Which is the Dykhne formula as required.

We next derive the non-adiabatic transition probabilities for the non-adiabatic entanglement transfer protocol. In the aim of deriving a simple expression, we adapt a constraint condition, namely that the pump, $g_3(t)$, and Stokes fields, $g_1(t)$, converge \cite{Vitanov1999}- such that the dynamical Stark shifts due to modes 1 and 2 cancel away from the avoided crossing- yielding constraint,

\begin{equation}
	t_{p2}=\sqrt{\abs{\left(\dfrac{t_s\tau_s}{\tau_{p2}}\right)^2-2\tau_{p2}^2\log\abs{\dfrac{A_{p2}}{A_{s}}}}}\label{tp3cond}
\end{equation}

In addition, we have required that the zero crossing of the diabatic energy $\tilde{\Delta}^{\text{eff}}_{2}(t)$ coincides with the local maximum of the effective coupling $g^{\text{eff}}_{3}(t)$, where $g^{\text{eff}}_{3}(t)$ is maximized at $t_{c}=\dfrac{t_{p3}\tau_{s}^2+t_{s}\tau_{p3}^2}{\tau_{s}^2+\tau_{p3}^2}$. We therefore obtain constraint,

\begin{equation}
	t_{\alpha}=\dfrac{t_{p3}\tau_{s}^2+t_{s}\tau_{p3}^2}{\tau_{s}^2+\tau_{p3}^2}+\dfrac{3\left(\abs{g_3(t_c)}^2-\abs{g_1(t_c)}^2\right)}{2\Delta_1\alpha_{0}}\label{tp3cond2}
\end{equation}

The conditions \eqref{tp3cond} and \eqref{tp3cond2}, while not strictly necessary, create a strong correspondence with the avoided crossing model seen in Fig. \ref{fig: LZmodel} and generate the small narrow energy gap at the avoided crossing for robust non-adiabatic population transfer, as seen in Figs. \ref{fig: phase1evo2} and \ref{fig: phase2evo2}. We will show that this leads to the natural result that the non-adiabatic transition probability is given by the Landau-Zener formula. The derivation below will lead us to explicit constraints that we can use to get a specific non-adiabatic transition probability.

We rescale time such that $i\tau_{p2}z=t^{\prime}-t_{p2}$. Using \eqref{FST}, we derive $A(z)$ for Hamiltonian $H_{1,2}^{\text{eff}}$, given by \eqref{THHam},
\begin{equation}
\begin{split}
    &A(z)=i\alpha_0\tau_{p2}^2\sqrt{A_1(z)^2-A_2(z)^2}\\
    &A_1(z)=Ce^{\frac{z^2}{2}\left(1+\left(\frac{\tau_{p2}}{\tau_s}\right)^2\right)}e^{i\frac{2z_{s}\tau_{p2}^2}{\tau_s^2}z}\\
    &A_2(z)=(z+iz_{\alpha})-\dfrac{i}{2}\left(\braket{rrg,000}{H_S\left(i\tau_{p2}z+t_{p2}\right)|rrg,000}\right.\\
        &\left.-\braket{ggg,101}{H_S\left(i\tau_{p2}z+t_{p2}\right)|ggg,101}\right)
	\end{split}
\end{equation}
Where $C=\dfrac{\sqrt{8}A_{p2}A_s}{\Delta_1\alpha_0\tau_{p2}}$. The turning points $\{\tilde{z}_{n,\mp}\}$ satisfying the equation $A(z)=0$, in the case where $A_2(z)\approx (z+iz_{\alpha})$ at the turning points, are given by,
\begin{equation}
	\tilde{z}_n=\mp i\sqrt{\dfrac{W_n\left(-2C^{2}\right)}{2}}\label{WZturn}
\end{equation}
Where $W_n(z)$ is the nth branch of the Lambert W function. If $2C^2\ll e^{-1}$, then the $n=0$ solution will dominantly contribute to the non-adiabatic transition term, given by Eq. \eqref{Dykhneres}. 

With $A(z)$ derived, we can calculate the phase integral in the Dykhne formula, where the contribution from the turning point $\tilde{z}_{0,+}$ is given by,
\begin{equation}
	\int_{0}^{\text{Re}\tilde{z}_{0,+}}d\eta\text{ }2A(\eta+i\text{Im}\tilde{z}_{0,+})\label{Azintff}
\end{equation}

Under the conditions \eqref{tp3cond} and \eqref{tp3cond2}, the imaginary part of the integrand in the integration domain in Eq. \eqref{Azintff} is approximately an elliptical function with semi-axes $2\text{Im}A(i\text{Im}\tilde{z}_{0,+})$ and $\text{Re}\tilde{z}_{0,+}$. 
We therefore arrive at the analytical result for the non-adiabatic transition probability,
\begin{equation}
	\begin{split}
		&\text{Pr}\approx\exp\left(-\pi\alpha_0(C\tau_{p3})^2(1+C^2)\right)\\
		&=\exp\left(-2\pi\dfrac{\left(\dfrac{2A_{p3}A_s}{\Delta_1}\right)^2}{\alpha_0}\right),
	\end{split}
\end{equation}
Which is the Landau-Zener formula. We find that although the non-adiabatic probability is independent of the pulse width $\tau_{p3}$, it is still required to satisfy condition $2C^2\ll e^{-1}$ to use this result- which requires a sufficiently large frequency sweep for the chirp over one pulse width, $\abs{\alpha_0\tau_{p3}}$. For adiabatic evolution, we require that the coupling rates are large enough to generate a sufficiently large energy gap at the avoided crossing.

\section{Derivation of the modes of the 3D rectangular cavity with a single moving mirror}\label{App5B}

The Klein-Gordon action, in a flat spacetime, describes the evolution of the electromagnetic field $\phi$. With external boundary conditions, set by the enclosing surface $\partial D$, we can determining the modes that lead to diagonal representation of the Hamiltonian. Our case is the 3D cavity with a uniformly moving mirror, with motion along the $x$ direction. The $y$ and $z$ boundaries ($y=0$, $y=L_y$, $z=0$, $z=L_z$) are fixed, where $\phi=0$. We have dynamic boundary conditions for the $x$-coordinate ($x=0$, $x=L_x(t)=L_{x,0}+vt$). Therefore the action is given by,

\begin{equation}
    S=\int dt\int_{0}^{L_x(t)} dx\int_0^{L_y} dy\int_0^{L_z} dz\text{ }\partial_{\mu}\phi\partial^{\mu}\phi\label{Act}
\end{equation}

We label the spatial integration domain above as $D(t)$. The modes can be found by solving the classical Klein-Gordon equation,

\begin{equation}
    \Box^2\phi(t,\vec{x})=0\label{KG}
\end{equation}

While we can find instantaneous plane wave modes with wavevector $\vec{k}=\left(\dfrac{\pi n_x}{L_x(t)}, \dfrac{\pi n_x}{L_y},\dfrac{\pi n_x}{L_z}\right)$ and positive/negative frequencies $\pm\omega_k=c\abs{\vec{k}}$ for which we can describe the result cavity dynamics and atom-cavity interaction terms, these will not diagnalize the Hamiltonian nor give fixed particle numbers assuming no interaction. Nonetheless we can use plane waves to describe the in- and out- modes when the mirror is stationary at times $t=0$ and $t=t_f$ respectively. The instantaneous creation/annihilation operations can then be found using canonical quantization of the cavity at a fixed time using the (positive frequency) modes given by,

\begin{equation}
    \begin{split}
        &\phi^{(\mu)}_{\vec{k},\text{ins}}(\vec{r},t)=\dfrac{e^{-i\omega_k(t)t}}{\sqrt{2\omega_{k}(t)}}\mathbf{u}^{(\mu)}_{\vec{k},\text{ins}}(\vec{r},t)\\
        &\mathbf{u}^{(\mu)}_{\vec{k},\text{ins}}(\vec{r},t)=\dfrac{\sqrt{8}\mathbf{e}^{(\mu)}_{\vec{k}}}{\sqrt{L_yL_zL_x(t)}}\sin(k_x(t) x)\sin(k_y y)\sin(k_z z)
    \end{split}\label{Basispl}
\end{equation}

The canonical momentum is given by $\Pi=\dfrac{\delta S}{\delta\phi}=\dot{\phi}$. The quantization relation $\comm{\phi(\vec{r},t)}{\Pi(\vec{r}^{\prime},t)}=i\delta(\vec{r}-\vec{r}^{\prime})$ is imposed. The Hamiltonian is given by,

\begin{equation}
    H(t)=\int_{D(t)} dV\text{ }\partial_{\mu}\phi\partial_{\mu}\phi\label{ActH}
\end{equation}

Giving us the Heisenberg equations,

\begin{equation}
    \begin{split}
        &\dot{\phi}(\vec{r},t)=i\comm{H(t)}{\phi(\vec{r},t)}=\Pi(\vec{r},t)\\
        &\dot{\Pi}(\vec{r},t)=i\comm{H(t)}{\Pi(\vec{r},t)}=\grad^2{\phi}(\vec{r},t)
    \end{split}\label{Heis}
\end{equation}

We calculate the Fourier transformed coordinate, $Q_{\vec{k},\text{ins}}$, and momentum, $P_{\vec{k},\text{ins}}$, in the instantaneous basis \eqref{Basispl},

\begin{equation}
    \begin{split}
        &Q^{(\mu)}_{\vec{k},\text{ins}}(t)=\int_{D(t)}dV\text{ }\phi(\vec{r},t)\mathbf{u}^{(\mu)}_{\vec{k},\text{ins}}(\vec{r},t)\\
        &P^{(\mu)}_{\vec{k},\text{ins}}(t)=\int_{D(t)}dV\text{ }\Pi(\vec{r},t)\mathbf{u}^{(\mu)}_{\vec{k},\text{ins}}(\vec{r},t)
    \end{split}
\end{equation}

Which allows us to define the instantaneous plane wave creation/annihilation operators,

\begin{equation}
    \begin{split}
        &a^{(\mu)}_{\vec{k},\text{ins}}(t)=\dfrac{\omega_{\vec{k}}(t)Q^{(\mu)}_{\vec{k},\text{ins}}(t)+iP^{(\mu)}_{\vec{k},\text{ins}}(t)}{\sqrt{2\omega_{\vec{k}}(t)}}\\
        &a^{(\mu)\dag}_{\vec{k},\text{ins}}(t)=\dfrac{\omega_{\vec{k}}(t)Q^{(\mu)}_{\vec{k},\text{ins}}(t)-iP^{(\mu)}_{\vec{k},\text{ins}}(t)}{\sqrt{2\omega_{\vec{k}}(t)}}\label{Insmods}
    \end{split}
\end{equation}

As previously mentioned, the instantaneous operators do not diagonalize the Hamiltonian. Following the approach by Law \cite{Law1994}, the Heisenberg equations of motion for these operators instead obey a different Hamiltonian $H_{\text{ins}}(t)$,

\begin{align}
    \begin{split}
        &H_{\text{ins}}(t)=\sum_{\vec{k},\mu}\omega_{\vec{k}}(t)a_{\vec{k},\text{ins}}^{(\mu)\dag}(t)a_{\vec{k},\text{ins}}^{(\mu)}(t)\\
        &+i\sum_{\vec{k},\mu}\chi_{\vec{k},\text{ins}}^{(\mu)}(t)\left(a_{\vec{k},\text{ins}}^{(\mu)\dag}(t)^2-a_{\vec{k},\text{ins}}^{(\mu)}(t)^2\right)\\
        &+\dfrac{i}{2}\sum_{\vec{k},\vec{k}^{\prime},\mu,\mu^{\prime}}\zeta_{\vec{k},\vec{k}^{\prime}}^{(\mu\mu^{\prime})}(t)\left(a_{\vec{k},\text{ins}}^{(\mu)\dag}(t)a_{\vec{k}^{\prime},\text{ins}}^{(\mu^{\prime})\dag}(t)\right.\\
        &\left.+a_{\vec{k},\text{ins}}^{(\mu)\dag}(t)a_{\vec{k}^{\prime},\text{ins}}^{(\mu^{\prime})}(t)-\text{h.c.}\right)
    \end{split}
\end{align}

where,

\begin{equation}
    \begin{split}
        &\chi_{\vec{k},\text{ins}}^{(\mu)}(t)=\dfrac{G_{\vec{k},\vec{k}}^{(\mu\mu)}(t)}{2}-\dfrac{1}{4}\dfrac{\partial}{\partial t}\log L_x(t)\\
        &\zeta_{\vec{k},\vec{k}^{\prime}}^{(\mu\mu^{\prime})}(t)=\sqrt{\dfrac{\omega_{\vec{k}}(t)}{\omega_{\vec{k}^{\prime}}(t)}}G_{\vec{k},\vec{k}^{\prime}}^{(\mu\mu^{\prime})}(t)\\
        &G_{\vec{k},\vec{k}^{\prime}}^{(\mu\mu^{\prime})}(t)=-\int_{D(t)}dV\text{ }\mathbf{u}^{(\mu)}_{\vec{k},\text{ins}}(\vec{r},t)\dfrac{\partial}{\partial t}\mathbf{u}^{(\mu^{\prime})}_{\vec{k}^{\prime},\text{ins}}(\vec{r},t)
    \end{split}
\end{equation}

Where the term $G_{\vec{k},\vec{k}^{\prime}}^{(\mu\mu^{\prime})}(t)$ is simplified to,

\begin{equation}
    \begin{split}
        &G_{\vec{k},\vec{k}^{\prime}}^{(\mu\mu^{\prime})}(t)=\left(\dfrac{\partial}{\partial t}\log(L_x(t))\right)\left(\dfrac{\delta_{\vec{k},\vec{k}^{\prime}}^{(\mu\mu^{\prime})}}{2}+\right.\\
        &\left.\dfrac{2k^{\prime}_x(t)}{L_x(t)}\int_0^{L_x(t)}dx\text{ }\sin(k_x(t) x)x\cos(k^{\prime}_x(t) x)\right)
    \end{split}
\end{equation}

In the instantaneous basis, the Hamiltonian reveals a Kerr non-linearity and inter-mode coupling term, with both terms' strength depending on the ratio of the instanteous mirror velocity and the instantaneous cavity length. Alternatively, this implies that minimizing the ratio will make the Kerr nonlinearity and the inter-mode coupling negligible.

The positive frequency in- and out- modes are given by,

\begin{align}
    \begin{split}
        &\phi^{(\mu)}_{\vec{k},\text{in}}(\vec{r},t)=\dfrac{e^{-i\omega_kt}}{\sqrt{2\omega_{k}}}\mathbf{u}^{(\mu)}_{\vec{k},\text{ins}}(\vec{r},0)\\
        &\phi^{(\mu)}_{\vec{k},\text{out}}(\vec{r},t)=\dfrac{e^{-i\omega_kt}}{\sqrt{2\omega_{k}}}\mathbf{u}^{(\mu)}_{\vec{k},\text{ins}}(\vec{r},t_f)\\
    \end{split}
\end{align}

where $\vec{k}$ satisfies the static boundary conditions at $t=0$, $t=t_f$ for the in- and out- modes respectively. The annihilation operators corresponding to these modes can be similarly defined using Eq. \eqref{Insmods}.

Continuity of the photon operators implies we can define $\hat{a}^{(\mu)}_{\vec{k},\text{in}}$, $\hat{a}^{(\mu)}_{\vec{k},\text{out}}$ in terms of the operators $\hat{a}^{(\mu)}_{\vec{k}}$ that generate the modes when the mirror is in motion.

The general field $\phi(\vec{r},t)$ for $t\in(0,t_f)$ can be expanded in terms of modes $\phi_{\vec{k}}$, to be determined, that solve Eq. \eqref{KG}. We loosely follow the quantization approach in \cite{Wald1995} in our following derivation,

\begin{equation}
    \begin{split}
        &\phi(t,\vec{r})=\sum_{\vec{k}}a_{\vec{k}}(t)\phi_{\vec{k}}(t,\vec{r})+a_{\vec{k}}^{\dag}(t)\phi^{\dag}_{\vec{k}}(t,\vec{r})\\
        &\phi_{\vec{k}}(t,\vec{r})=T(x,t)\mathbf{u}^{(\mu)}_{\vec{k}}(\vec{r},t)\\
        &\mathbf{u}^{(\mu)}_{\vec{k}}(\vec{r},t)=X(x,t)Y(y)Z(z)
    \end{split}
\end{equation}

The Klein-Gordon inner product must be satisfied,

\begin{eqnarray}
    &\left(\phi_{\vec{k}},\phi_{\vec{k}^{\prime}}\right)=\int_{D(t)}dV\text{ }\phi_{\vec{k}}(\vec{r},t)^{*}\overleftrightarrow{\partial_t}\phi_{\vec{k}^{\prime}}(\vec{r},t)
\end{eqnarray}

From which we can define the commutation relation of the operator $\hat{\Omega}[\phi_{\vec{k}},\cdot]=ia(\overline{K\phi}_{\vec{k}})-ia^{\dag}(K\phi_{\vec{k}})$, where $a(\overline{K\phi}_{\vec{k}})$ is the annihilation operator associated with the positive frequency part of $\phi_{\vec{k}}$,

\begin{equation}
    \begin{split}
        &\comm{\hat{\Omega}[\phi_{\vec{k}},\cdot]}{\hat{\Omega}[\phi_{\vec{k}^{\prime}},\cdot]}=-i\left(\phi_{\vec{k}},\phi_{\vec{k}^{\prime}}\right)\label{QuaKG}
    \end{split}
\end{equation}

The positive and negative frequency parts, $\phi_{\vec{k}}^{\pm}(\vec{r},t)$, corresponding to $\phi_{\vec{k}}(\vec{r},t)$, satisfy the equation,

\begin{equation}
    \dot{\phi}_{\vec{k}}^{\pm}(\vec{r},t)=-i\omega_{\vec{k},\pm}(t)\phi_{\vec{k}}^{\pm}(\vec{r},t)
\end{equation}

where $\text{Re}\omega_{\vec{k},\pm}(t)>0, <0$ for positive/negative frequency modes respectively.

To derive the form of the classical Klein-Gordon modes, we move to a coordinate frame where the boundary conditions are static using transformation,

\begin{equation}
    \begin{split}
        &\tau=\sqrt{\left(\dfrac{L_{x,0}}{v}+t\right)^2+\left(\dfrac{x}{c}\right)}\\
        &\xi=\sinh^{-1}\left(\dfrac{x}{c\tau}\right)
    \end{split}
\end{equation}

The dynamic boundary conditions in $x$ are now given by $\xi=0$, $\xi=\xi_1=\sinh^{-1}\left(v/c\left(1-\left(v/c\right)^2\right)^{-1/2}\right)$. And Eq. \eqref{KG} has the below form,

\begin{equation}
    \dfrac{\partial^2\phi}{\partial\tau^2}+\dfrac{1}{\tau}\dfrac{\partial\phi}{\partial\tau}-\dfrac{1}{\tau^2}\dfrac{\partial^2\phi}{\partial\xi^2}-\left(\dfrac{\partial^2\phi}{\partial y^2}+\dfrac{\partial^2\phi}{\partial z^2}\right)=0
\end{equation}

Using separation of variables, we solve for $\phi=T(\tau)X(\xi)Y(y)Z(z)$ where,

\begin{equation}
    \begin{split}
        &Y(y)=\sqrt{\dfrac{2}{L_y}}\sin \left(k_y y\right)\\
        &Z(z)=\sqrt{\dfrac{2}{L_z}}\sin \left(k_z z\right)\\
        &X(\xi)=\sqrt{\dfrac{2}{\xi_1}}\sin \left(k_{\xi}\xi\right),	
    \end{split}
\end{equation}

Where $k_y=\dfrac{\pi n_y}{L_y}$, $k_z=\dfrac{\pi n_z}{L_z}$, $k_{\xi}=\dfrac{\pi n_{\xi}}{\xi_1}$ and $\kappa=\sqrt{k_y^2+k_z^2}$ and $T(\tau)$ satisfies the Bessel equation,

\begin{equation}
    \begin{split}
        \dfrac{\partial^2\phi}{\partial\tau^2}+\dfrac{1}{\tau}\dfrac{\partial\phi}{\partial\tau}+\left(\kappa^2-\dfrac{\left(ik_{\xi}\right)^2}{\tau^2}\right)=0
    \end{split}
\end{equation}

Which has different solutions depending on the value of $\kappa$. If $\kappa=0$,

\begin{equation}
    \begin{split}
        &T(\tau)=c_1\left(\dfrac{v\tau}{L_{x,0}}\right)^{ik_{\xi}}+c_2\left(\dfrac{v\tau}{L_{x,0}}\right)^{-ik_{\xi}}
    \end{split}
\end{equation}

If $\kappa\neq 0$,

\begin{equation}
    \begin{split}
        &T(\tau)=\Gamma(1-ik_{\xi})\left(c_1J_{ik_{\xi}}(\kappa\tau)+c_2Y_{ik_{\xi}}(\kappa\tau)\right)\\
        &+\Gamma(1+ik_{\xi})\left(c_3J_{-ik_{\xi}}(\kappa\tau)+c_4Y_{-ik_{\xi}}(\kappa\tau)\right),
    \end{split}
\end{equation}

where $J_{i\nu}(z)$, $Y_{i\nu}(z)$ is the Bessel function of the first/second kind with imaginary order. We  normalize our obtained modes, with mode vector $\vec{n}$ using the Klein-Gordon inner product,

\begin{equation}
    -i\tau\int_{0}^{\xi_1} d\xi \int_0^{L_y} dy\int_0^{L_z} dz\text{ } \phi^{*}_{\vec{k}^{\prime}}\overleftrightarrow{\partial_{\tau}}\phi_{\vec{k}}=\delta_{\vec{k}^{\prime},\vec{k}}
\end{equation}

We can revert to the original $(t,x,y,z)$ coordinates to obtain the stationary modes for the system described by the action \eqref{Act}. 

In general, it's not a simple exercise to separate the classical modes into positive and negative frequency components due to the general non-separability of $\phi_{\vec{k}}^{\pm}(\vec{r},t)$ into a product of functions that depend exclusively on $t$ or $x$. However this is not a worry if relativistic effects can be ignored.

In the non-relativistic case $v\ll c$ along with the assumption that the maximum change in the cavity dimensions is much smaller than the original dimension $vT\ll L_{x,0}$, we obtain the below approximations,

\begin{equation}
    \begin{split}
        &\tau\approx\dfrac{L_{x,0}}{v}+t\\
        &\xi\approx\dfrac{vx}{cL_x(t)}\\
        &\xi_1\approx\dfrac{v}{c},
    \end{split}
\end{equation}

This approximates $X(\xi)$ to,

\begin{equation}
    \begin{split}
        &X(\xi)\approx\sqrt{\dfrac{2c}{v}}\sin\left(k_x(t)x\right)
    \end{split}
\end{equation}

where $k_x(t)=\dfrac{\pi n_{\xi}}{L_x(t)}$

For $\kappa=0$, $T(\tau)$ approximates,

\begin{equation}
    \begin{split}
        &T(\tau)\approx c_1\left(1+\dfrac{vt}{L_{x,0}}\right)^{ick_x(t)(\frac{L_{x,0}}{v}+t)}\\
        &+c_2\left(1+\dfrac{vt}{L_{x,0}}\right)^{-ick_x(t)(\frac{L_{x,0}}{v}+t)}
    \end{split}
\end{equation}

And using the limit $\lim_{n\rightarrow+\infty}\left(1-\dfrac{x}{n}\right)^{in}=e^{-ix}$, we further simplify to,

\begin{equation}
    \begin{split}
        &T(\tau)\approx c_1e^{ick_x(t)t}\left(1+\dfrac{vt}{L_{x,0}}\right)^{-ick_x(t)t}\\
        &+c_2e^{-ick_x(t)t}\left(1+\dfrac{vt}{L_{x,0}}\right)^{ick_x(t)t}
    \end{split}
\end{equation}

For $\kappa\neq 0$, approximating is significantly more difficult due to the complexity in dealing with Bessel functions of large imaginary order and large argument. We use a result \cite{Watson1995} derived using the stationary phase approximation. For the negative frequency solution, we have,

\begin{align}
    \begin{split}
        &T_{-}(t)=J_{-ik_{\xi}}(\kappa t)\approx\dfrac{e^{ik_{\xi}\left(\tanh\left(\gamma\right)-\gamma\right)}e^{i\pi/4+k_{\xi}\pi/2}}{2\sqrt{-i\pi k_{\xi}\tanh\left(\gamma\right)}},
    \end{split}
\end{align}

where $\cosh(\gamma)=\dfrac{ik_{\xi}}{\kappa(L/v+t)}$. With some algebraic manipulations, we obtain the final result,

\begin{align}
    \begin{split}
        &T_{-}(t)\propto\dfrac{e^{i\sqrt{\kappa^2+k_x(t)}t}}{\sqrt{\kappa^2+k_x(t)^2}L_x(t)/v}
    \end{split}
\end{align}

Putting everything together, we have,

\begin{equation}
    \begin{split}
        &\phi\propto\sqrt{\dfrac{8}{L_yL_zL_x(t)}}\sin \left(k_x x\right)\sin \left(k_y y\right)\sin \left(k_z z\right)\\
        &\times\dfrac{e^{i\sqrt{\kappa^2+k_x(t)}t}}{\sqrt{(\kappa^2+k_x(t)^2)L_x(t)}}
    \end{split}	
\end{equation}

The above term is normalized by multiplying by factor $\left(\left(\kappa^2+k_x(t)\right)L_x(t)^2\right)^{1/4}$, giving us the relation $\phi_{\vec{k}}^{(\mu)\dag}(t,\vec{r})\approx\phi_{\vec{k},\text{ins}}^{(\mu)\dag}(t,\vec{r})$.

Hence, we can make the following identification for the positive frequency mode,

\begin{align}
    \begin{split}
        &{\phi}_{\vec{k}}^{(\mu)+}(\vec{r},t)\sim(J_{-ik_{\xi}}(\kappa t))^{*}
    \end{split}
\end{align}

And associate the creation operator $a_{\vec{k}}^{(\mu)\dag}$ accordingly with the normalized mode ${\phi}_{\vec{k}}^{(\mu)+}(t)$. The Hamiltonian is diagonal in this representation with $H(t)=\sum_{\vec{k},\mu}\omega_{k,+}(t)a_{\vec{k}}^{(\mu)\dag}a_{\vec{k}}^{(\mu)}$. We derive the dipole approximation interaction Hamiltonian, where the cavity is coupled to a two-level atom located at $\vec{r}=\vec{r}_0$,

\begin{align}
    \begin{split}
        &H_{\text{int}}(t)=\sum_{\vec{k},\mu}(g_{\vec{k}}^{(\mu)}(t)a_{\vec{k}}^{(\mu)}\sigma^{+}+\text{H.c.})\\
        &g_{\vec{k}}^{(\mu)}(t)=\dfrac{\omega_{k,+}(t)}{\sqrt{2\hbar\epsilon_0}}{\phi}_{\vec{k}}^{(\mu)+}(\vec{r}_0,t)\expval{2|\vec{d}^{+}\cdot\mathbf{e}_{\vec{k}}^{(\mu)}|1}
    \end{split}
\end{align}

We obtain the result that the 3D cavity with a moving mirror can be described using the instantaneous cavity modes.

\bibliography{bibmirrorcav}

\end{document}